\documentclass[aps,prb,twocolumn,superscriptaddress,longbibliography]{revtex4-2}

\usepackage[utf8]{inputenc}
\usepackage{physics}
\usepackage{comment}
\usepackage{amssymb}
\usepackage[colorlinks=true,linkcolor=MidnightBlue,citecolor=RedViolet,urlcolor=BlueViolet]{hyperref} 
\usepackage{graphicx}
\usepackage{amsmath}
\usepackage{dcolumn}
\usepackage{bm}
\usepackage{graphicx}
\usepackage{amsmath}
\usepackage{amsfonts}
\usepackage{color}
\usepackage{mathtools}
\usepackage{braket}
\usepackage{gensymb}
\usepackage{textcomp}
\usepackage{soul}
\usepackage[dvipsnames]{xcolor}
\usepackage{enumitem}

\usepackage[normalem]{ulem}
\newcommand{\editor}[2]{%
  \expandafter\newcommand\csname #1note\endcsname[1]{%
    \textcolor{#2}{[\textbf{#1:}  ##1]}}%
  \expandafter\newcommand\csname #1\endcsname[1]{%
    \textcolor{#2}{##1}}%
  \expandafter\newcommand\csname #1cancel\endcsname[1]{%
    \textcolor{#2}{\sout{##1}}}%
  \expandafter\newcommand\csname #1change\endcsname[2]{%
    \textcolor{#2}{\sout{##1} ##2}}%
  \newenvironment{#1text}{\color{#2}}{\color{black}}
}
\editor{AB}{NavyBlue}
\editor{MG}{teal}
\editor{AM}{red}

\begin{document}
\title{\textcolor{black}{Symmetry-Guided Computational Screening of Two-Dimensional Altermagnets with \textit{ab initio} Hubbard Corrections}}



\author{Anumita Bose}
\email[]{abose@sissa.it}
\affiliation{Scuola Internazionale Superiore di Studi Avanzati (SISSA), I-34136 Trieste, Italy}

\author{Nataliia Manko}
\affiliation{Scuola Internazionale Superiore di Studi Avanzati (SISSA), I-34136 Trieste, Italy}

\author{Marco Gibertini}
\email[]{marco.gibertini@unimore.it}
\affiliation{Dipartimento di Scienze Fisiche, Informatiche e Matematiche, Università degli Studi di Modena e Reggio Emilia, Via G. Campi 213/a, 41125 Modena, Italy}
\affiliation{Istituto Nanoscienze -- CNR, S3, 41125 Modena, Italy}

\author{Antimo Marrazzo}
\email[]{amarrazz@sissa.it}
\affiliation{Scuola Internazionale Superiore di Studi Avanzati (SISSA), I-34136 Trieste, Italy}

\date{\today}

\begin{abstract}
Altermagnets combine compensated antiferromagnetic order with momentum-dependent spin splitting, offering a promising platform for spintronic applications without macroscopic magnetization or stray magnetic fields. Although a wide range of three-dimensional (3D) materials have been identified as altermagnets, two-dimensional (2D) altermagnets remain comparatively limited. In this work, we perform a high-throughput computational search for altermagnetism across 2710 materials in the Materials Cloud 2D Crystals (MC2D) database. Our approach combines symmetry-based screening with first-principles density functional theory calculations, including self-consistent Hubbard-$U$ corrections, to reliably capture magnetic ground states. Through a systematic exploration of magnetic configurations and their energetic stability, we identify 42 materials exhibiting altermagnetic ground states for at least one value of $U$, of which 24 remain robust upon determination of the Hubbard-$U$ parameters from first principles\textemdash including 4 materials previously reported in the literature and 20 newly predicted candidates. These comprise promising monolayers such as metallic Fe$_2$Si$_2$SbO$_9$, and insulating CoBrO, with spin splittings about 294 meV and 330 meV, respectively. Our results significantly expand the pool of potential 2D altermagnet candidates with favorable exfoliation energetics and provide valuable guidance for experimental efforts. In addition, this work establishes a high-throughput computational framework for reproducible discovery and characterization of altermagnetic materials.

\end{abstract}

\maketitle

Altermagnetism is a recently identified magnetic phase that occupies a distinct position within the broader landscape of magnetic order. Altermagnets combine traits of both ferromagnets (FMs) and conventional collinear antiferromagnets (AFMs)\textemdash possessing a collinear AFM-like magnetic order in real space with no net magnetization, and time reversal symmetry (TRS) broken spin-splitting in $k$-space, akin to FMs~\cite{vsmejkal2022emerging, vsmejkal2022beyond, bai2024altermagnetism, krempasky2024altermagnetic, fender2025altermagnetism, song2025altermagnets, tamang2025altermagnetism}. The momentum dependent spin-split band structure of altermagnets enables independent control of spin channels unlike AFMs, while avoiding the drawbacks of macroscopic magnetization and stray magnetic fields that are commonly associated with FMs.
As a result, altermagnets provide a unique platform for spin-dependent transport phenomena and offer promising opportunities for the realization of energy-efficient spintronic functionalities. These prospects have motivated extensive theoretical and experimental efforts aimed at understanding their fundamental properties and exploring their potential for next-generation spintronic technologies~\cite{vsmejkal2022giant,gonzalez2021efficient,karube2022observation,bai2022observation,gonzalez2024anisotropic,chen2024emerging,song2025altermagnets,zhang2025theory,guo2025spin}.

Since the discovery of two-dimensional (2D) magnetic order in van der Waals systems~\cite{lee2016ising, gong2017discovery,huang2017layer,burch2018magnetism}, 2D magnets have propelled rapid progress in modern spintronics~\cite{gong2019two,li2019intrinsic,khan2020recent,liu2023emergent}. This excitement is further fueled by the inherent advantages of 2D materials, such as high sensitivity to external stimuli, efficient controllability of their physical properties, and ease of stacking into heterostructures with diverse functionalities~\cite{novoselov20162d, liu2016van, chen2019electrically, du2021engineering, mak2019probing}. Consequently, exploring 2D altermagnetic systems offers an exciting frontier in spintronics with the potential for reduced, faster, and more energy efficient devices.

Although a wide range of three-dimensional (3D) altermagnets have been investigated~\cite{vsmejkal2022beyond,guo2023spin, amin2024nanoscale,fedchenko2024observation,guo2024direct,reimers2024direct,ding2024large,gao2025ai,bhattarai2025high,sufyan2026high}, reports of intrinsic 2D altermagnets remain relatively limited, and their experimental realization continues to pose a significant challenge~\cite{ma2021multifunctional,jiang2025metallic,zhang2025crystal,zeng2026classification}. To circumvent this, extrinsic routes to 2D altermagnetism have been explored, including electric-field-induced spin splitting~\cite{mazin2023induced,wang2024electric}, gate-voltage control via spin-layer coupling~\cite{zhang2024predictable}, twisted van der Waals bilayers~\cite{nonrelativistic_He2023,liu2024twisted,nonrelativistic_Sheoran2024}, and bilayer stacking strategies as a general 
approach to engineering 2D altermagnetic order~\cite{pan2024general,zeng2024bilayer}. Xu et al. proposed a systematic chemical design strategy for discovering monolayer altermagnets, using symmetry-preserving structural modifications~\cite{xu2026chemical}. While these approaches offer versatile handles to engineer altermagnetic order, identifying materials with intrinsic 2D altermagnetism remains the more fundamental pursuit. 
On the theoretical side, considerable efforts have been devoted to this end. Complementary symmetry-based studies have established spin layer group frameworks for classifying 2D altermagnetic materials~\cite{zeng2024description}, while pentagonal lattices have emerged as a particularly promising structural motif~\cite{wang2025pentagonal}.
Sødequist and Olsen performed a high-throughput screening of the C2DB database, identifying several structurally stable 2D altermagnets and characterizing their spin-orbit and magnonic properties~\cite{sodequist2024two}. More recently, Sufyan et al. extended the search to the MAGNDATA database~\cite{guo2023spin}, uncovering 180 candidate altermagnets with robust momentum-dependent spin splitting across bulk and 2D systems~\cite{sufyan2026high}. Notably, Haddadi et al. systematically explored the magnetic landscape of 194 easily exfoliable monolayers from the Materials Cloud 2D Crystals database (MC2D) database and identified 2 altermagnetic monolayers among them~\cite{haddadi2026exploring}.

Despite this rapid progress, existing high-throughput studies share a common methodological limitation: magnetic ground states are determined first and symmetry analysis is applied only afterward, with an unavoidable constraint to small unit cells to limit the computational cost~\cite{torelli2020high, sodequist2024two,haddadi2026exploring}. This approach inherently restricts the search space and risks overlooking a broad class of latent altermagnets whose crystal symmetry already permits spin splitting but whose magnetic ordering has not been systematically resolved. Motivated by this gap, in this work we adopt a symmetry-first strategy: we perform a symmetry-guided screening of the easily exfoliable magnetic monolayers in the  MC2D database~\cite{mounet2018two,campi2023expansion} to identify prospective altermagnets before any magnetic ground-state calculations are carried out. The shortlisted candidates are subsequently examined using Hubbard-corrected density-functional theory (DFT) calculations~\cite{liechtenstein1995density,cococcioni2005linear} for a range of Hubbard-$U$ values to establish potential ground-state altermagnets. At this stage we obtain 42 monolayers that have an altermagnetic  ground state for at least one $U$ value. For these selected compounds, the Hubbard-$U$ parameters are further refined self-consistently using density functional perturbation theory (DFPT)~\cite{baroni1987green,timrov2018hubbard,timrov2022hp}, targeting accurate predictions of energetic stability, electronic structure, and spin splitting. This approach is both computationally efficient and  enables exploration of a significantly broader material space. We identify 24 robust 2D altermagnetic candidates among easily exfoliable compounds, of which 4 have been previously reported in the literature. Beyond a curated database of candidate materials and their key properties, we provide a fully automated AiiDA-based workflow~\cite{huber2020aiida, uhrin2021workflows} that enables reproducible and scalable exploration of altermagnetism across broader materials spaces.


\begin{figure}
\centerline{\includegraphics[scale=0.4]{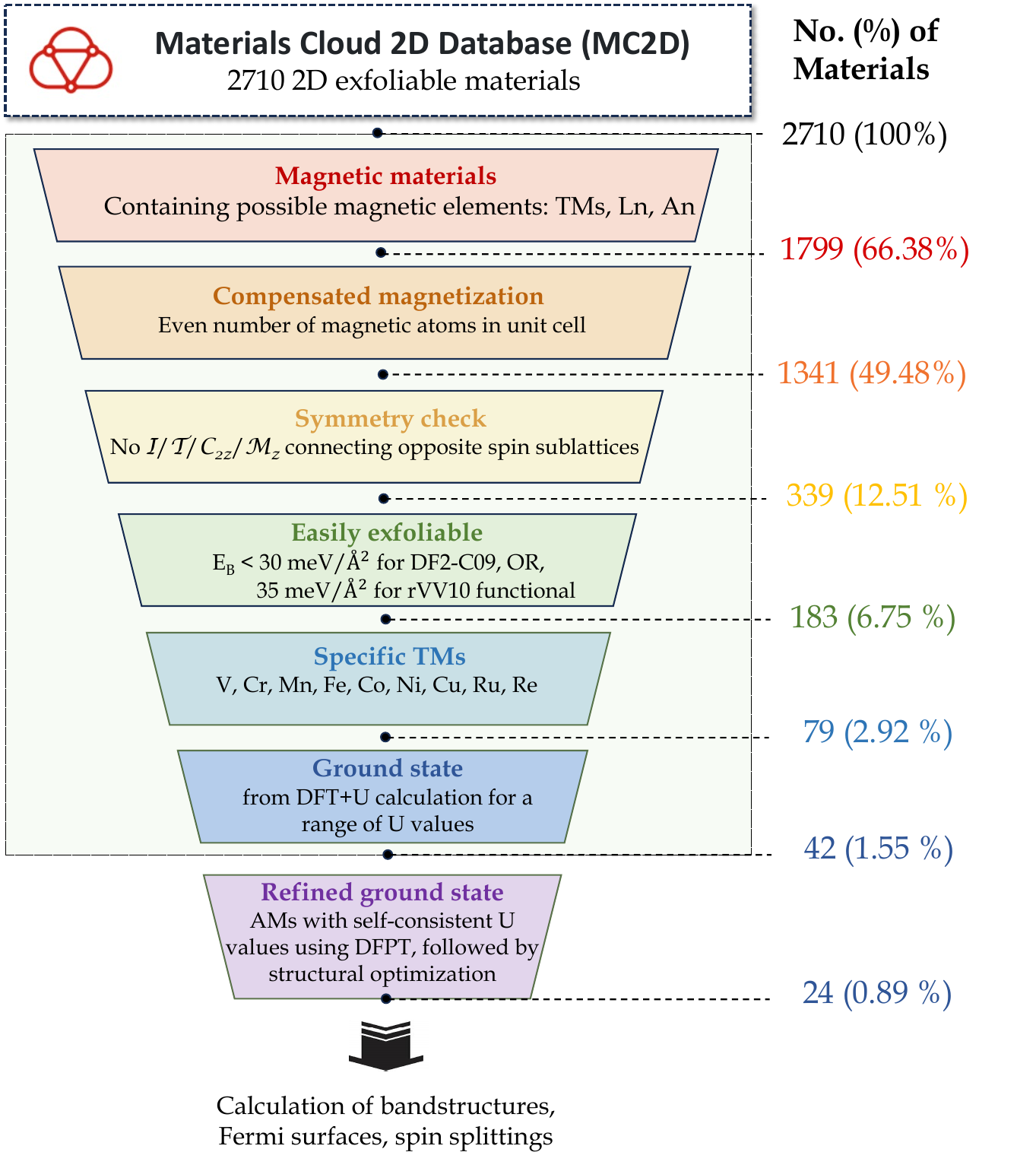}}
\caption{\textbf{Screening protocol for 2D altermagnets.} Schematic of the computational screening procedure employed to identify 2D altermagnets that can be exfoliated from experimentally known crystalline materials. Starting with a pool of 2710 2D compounds from the MC2D database~\cite{mounet2018two,campi2023expansion}, the number of candidates is successively narrowed down through a series of steps each shown in different colored zones, with the number of systems at each stage indicated. The overall high-throughput screening phase, shown within the light green shaded region, is followed by a low-throughput refinement phase. In the refinement phase, self-consistent Hubbard-$U$ parameters are first obtained using density functional perturbation theory (DFPT), followed by structural relaxations and DFT+$U$ calculations to evaluate the electronic band structures and altermagnetic spin splittings in the selected materials, yielding 24 refined ground state altermagnets.}
\label{HT2D}
\end{figure}

\begin{figure}
\centerline{\includegraphics[scale=0.45]{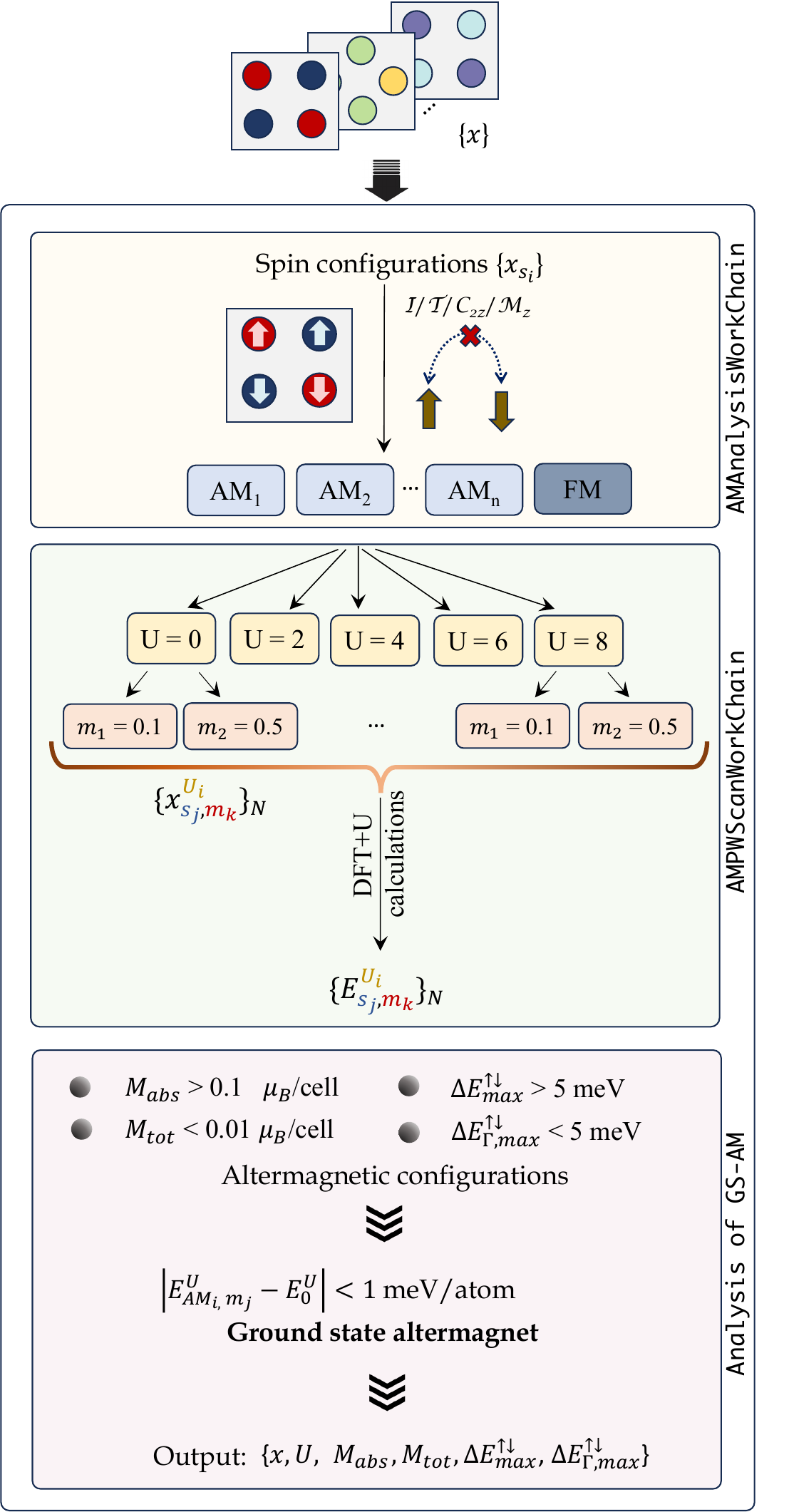}}
\caption{\textbf{Schematic representation of the automated high-throughput screening protocol.}  The AiiDA WorkChain is divided into three parts: 1. \texttt{AMAnalysisWorkChain}, which takes the crystal structures as input, and generates possible altermagnetic spin configurations based on symmetry analysis; 2. \texttt{AMPWScanWorkChain}, in which DFT calculations are performed for all possible altermagnetic configurations along with ferromagnetic one, considering a varying set of Hubbard-$U$ parameters and starting magnetization values; and 3. \texttt{Analysis of GS-AM}, where results are analyzed based on conditions mentioned in the figure, to identify the ground-state altermagnets (GS-AMs). \textcolor{black}{Relevant quantities, such as the total magnetization ($M_{tot}$), absolute magnetization ($M_{abs}$), maximum spin splitting ($\Delta E^{\uparrow\downarrow}_{max}$), and maximum spin splitting at $\Gamma$ ($\Delta E^{\uparrow\downarrow}_{\Gamma,max}$), are then tabulated in the output along with the input structure and the corresponding Hubbard-$U$ value.}}
\label{workflow}
\end{figure}


\section{Results and Discussion}
The screening process encompasses two stages. First we describe the high-throughput screening framework employed to identify candidate altermagnets from the MC2D database. Then we discuss the refined first-principles calculations with self-consistently determined Hubbard-$U$ parameters to validate the altermagnetic ground states and accurately evaluate their electronic structures and spin splittings. A schematic overview of the complete screening workflow is shown in Fig.~\ref{HT2D}. The light-green shaded block denotes the high-throughput stage, where successive filtering steps are indicated by different colors. The number and percentage of surviving candidates are reported after each stage of the workflow. The lower part of the schematic summarizes the subsequent fine-grained analysis used to obtain \textcolor{black}{the final predictions for material properties.}

\subsection{High-throughput screening}

In search of 2D altermagnets, we comprehensively screen 2,710 monolayer materials derived from experimentally known 3D parent compounds. These materials, available in the MC2D database, were identified through large-scale van der Waals density functional theory (vdW-DFT) calculations~\cite{mounet2018two,campi2023expansion}. All structures are geometrically optimized as isolated monolayers within the PBE approximation~\cite{perdew1996generalized} and classified as either easily exfoliable or potentially exfoliable from their bulk counterparts. To identify candidate 2D altermagnets, we apply the screening protocol outlined below.

As an initial filtering step, we retain only potentially magnetic materials. To this end, we consider only materials with at least one transition-metal, lanthanide, or actinide element, as their partially filled $d$- or $f$-shells can support finite magnetic moments. This criterion reduces the dataset to 1,799 materials. Since altermagnets require compensated magnetization within the unit cell, we retain structures in which at least one magnetic species is present with an even number of atoms, yielding 1,341 candidate materials. Having this set of materials, we next impose symmetry constraints relevant to 2D altermagnetism. Denoting with $E$, $\mathcal{C}_2$, $\mathcal{P}$, $\mathcal{T}$, $\tau$ respectively identity, spin-flip rotation, space inversion, TRS and lattice translation, in 3D magnetic materials, spin-symmetry operations such as \([E|| \mathcal{P}]\)\([\mathcal{C}_2 || \mathcal{T}] \equiv \mathcal{P}\mathcal{T}\) and \([\mathcal{C}_2 || \tau]\) transform an energy eigenstate \(E(k_\parallel, s)\) into \(E(k_\parallel, -s)\), thereby enforcing spin degeneracies across the 3D BZ. In two dimensions, however, additional symmetries involving out-of-plane operations, such as the combined symmetry \([E || \mathcal{C}_{2z}][\mathcal{C}_2 || \mathcal{T}]\) or \([\mathcal{C}_2 || \mathcal{M}_z]\), also cause band degeneracy throughout the 2D BZ~\cite{zeng2024description,sodequist2024two} ($\mathcal{C}_{2z}$ is the two-fold rotation about the vertical $z$-axis and $\mathcal{M}_z$ is the mirror symmetry with respect to the horizontal $xy$-plane). We therefore select materials where the magnetic ions can be divided into two  opposite spin sublattices~\cite{peng2025multicomponent}, that are connected by some crystal symmetry to enforce zero net magnetization, but exclude those for which the connection happens through $\mathcal{P}$, $\tau$, $\mathcal{C}_{2z}$ or $\mathcal{M}_z$. This procedure reduces the candidate set to 339 compounds. To further identify experimentally more promising systems, we retain only materials with favorable exfoliation energetics, using binding-energy thresholds of 30 meV \AA$^{-2}$ and 35 meV \AA$^{-2}$, calculated using the DF2-C09 and rVV10 functionals, respectively, which correspond to the ``easily exfoliable" criteria introduced in Ref.~\cite{mounet2018two}. This step yields 183 easily-exfoliable monolayers as potential 2D altermagnets. Finally, we focus on compounds containing transition metals commonly associated with robust magnetic ordering\textemdash namely V, Cr, Mn, Fe, Co, Ni, Ru, and Re\textemdash resulting in a set of 79 candidate materials. We note that these filtering steps are conservative, and the true number of 2D altermagnets in the database is likely larger than the 79 candidates identified here.

\begin{table*}[!t]
\centering
\caption{\textbf{2D altermagnetic candidates and their properties.} List of 2D altermagnetic candidates along with their corresponding space groups, calculated self-consistent Hubbard-$U$ values, and key properties derived from the DFT+$U$ calculations, such as, spin splitting values, band gap and energetic stability. The energetic stability, $\Delta E^{\mathrm{min}}_{\mathrm{AM-NAM}}$, is defined as the energy difference between the altermagnetic ground state and the lowest-energy non-altermagnetic magnetic state, evaluated for the unrelaxed crystal structure. The table further enlists range of Hubbard-$U$ parameters for which the material sustains altermagnetic behavior, as determined from the high-throughput screening.}
\label{Table_details}
\begin{tabular*}{\textwidth}{@{\extracolsep{\fill}} l c c c c c c c}
\hline\hline
\shortstack{Chemical \\ formula of \\ 2D AMs} & \shortstack{Space\\ group \\ (No.)} & \shortstack{Self consistent\\Hubbard-$U$\\ (eV)} & \shortstack{Maximum \\splitting \\ (meV)} &\shortstack{Maximum \\splitting near\\ E$_F$(meV)}& \shortstack{Band \\gap \\(eV)}  & \shortstack{Energetic \\stability ($\Delta E^{min}_{AM-NAM}$ in\\ meV/magnetic atom)} & \shortstack{AM for\\ $U$ values \\(eV)}\\
\hline
\hline



Fe$_2$SeTe & P4mm (99) &  5.3 & 487 & 236 &2.42  & 220 & 0, 2, 4, 6\\

CoBrO & P2$_1$/c (14)  & 5.9 & 330 & 66 & 1.31  &117 & 2, 4, 6, 8\\

CuH$_8$N$_2$(OF$_2$)$_2$ & P2$_1$/c (14)  & 5.2 & 317 &28&2.84 &-3 & 4, 6, 8\\ 

Fe$_2$Si$_2$SbO$_9$ & Cm (8) &  6.3  & 294 &292 &0.0&22 & 0, 4, 6, 8\\

NiH$_2$C$_2$N$_3$Cl & P2$_1$/c (14) & 7.9 & 290  & 4 &3.21 & 4 & 6, 8\\

SrCo$_2$Te$_3$(ClO$_4$)$_2$ & P2$_1$ (4)  & 6.4 & 251 &74& 2.96  & -5 & 6, 8\\

MnH$_2$C$_2$N$_3$Cl & P2$_1$/c (14) & 5.4 & 222  & 10 & 3.98  & 14 & 0, 2, 4, 6, 8\\

CoMoTeO$_{6}$   & P2$_1$2$_1$2 (18) &  6.7  &205 &89 & 3.13 & 11 & 0, 2, 4, 6, 8\\


MnMoTeO$_6$ & P2$_1$2$_1$2 (18) &  6.5  & 189 & 132 & 3.01  & 3 & 0, 2, 4, 6, 8\\

CoH$_8$N$_2$(OF$_2$)$_2$ & P2$_1$/c (14)  & 5.6 
& 186 &35 & 5.03 &28& 4, 6, 8 \\

Mn$_4$SnS$_4$ & Pmmm (47)  & 5.4 & 178 & 73 & 0.28 & 26 & 2, 4, 6, 8 \\

RuF$_4$ & P2$_1$/c (14) & 3.7 & 171 &49&1.67 &119 & 0, 2, 4, 6, 8 \\

ReCl$_{3}$ & C2 (5)   & 3.9 & 155 &67 &0.40 &242& 2, 4, 6, 8 \\

CuSe$_2$O$_5$ & Pmma (51)   & 6.7 & 133 &48& 2.05  &0 & 0, 2, 4, 6, 8\\

CuP$_2$(HO)$_4$ & P2$_1$/c (14)  & 5.9 
 & 117 &33&2.36 &0& 0, 2, 4, 6, 8\\


NiPH$_4$ClO$_3$ & P2$_1$/c (14)  & 7.3 & 96 & 31& 4.42&5&2, 4, 6, 8 \\

V$_4$S$_9$Br$_4$  & P2$_1$ (4)   & 4.9   & 91 &84& 0.0 &16& 0, 2, 4, 6, 8\\

V$_3$(H$_3$O$_5$)$_2$ & P4bm (100)  & 5.6 & 91 &21& 2.70  &-1& 2, 4 , 6, 8\\


CoH$_4$(NO$_4$)$_2$    & P2$_1$/c (14)  &  6.7  & 76&21 &3.58&0& 2, 4, 6 \\

AgRuF$_7$ & P2$_1$/c (14) & 3.8 & 74 &31&0.03 &2& 0, 2, 4, 6, 8\\

Al$_2$CuCl$_8$ & P2$_1$/c (14)&  6.5  & 66&5 &1.11&-4& 0, 2, 4, 6, 8 \\

CuH$_4$C$_6$S(NCl)$_2$ & P2$_1$/m (11) &  6.6 
 & 53 &9&1.58&0& 0, 2, 4, 6, 8 \\

CuH$_6$(CN$_2$)$_4$ & P2$_1$/c (14)   &  6.3  & 51 &13&2.03 &0& 0, 2, 4, 6, 8\\

VF$_4$ & P2$_1$/c (14) & 5.4 
& 27 &6&3.41 & 14& 0, 6\\

\hline \hline
\end{tabular*}

\end{table*}

Up to this stage, the screening procedure relies exclusively on chemical and symmetry-based filtering criteria. We next employ first-principles calculations to identify compounds capable of stabilizing altermagnetic ground states. Because the candidate materials contain partially filled $d$-shell elements, all calculations are performed within the DFT+$U$ framework to account for strongly localized $d$-electrons. Determining the Hubbard-$U$ parameter self-consistently using DFPT for the entire dataset is computationally demanding, hence we perform DFT+$U$ calculation across a considerable range of $U$ values, from 0 to 8 eV in increments of 2 eV. In addition to reducing the computational cost, the strategy provides an indirect metric to assess the robustness of the altermagnetic ground state and the reliability of the corresponding prediction. To mitigate the risk of being trapped in local minima of the total energy, each calculation is initialized using two distinct fractional magnetic moments ($m_i$ = 0.1 and 0.5) on the transition-metal sites. In addition to the symmetry-derived altermagnetic configurations, ferromagnetic states are also considered for comparison. We impose a set of quantitative criteria to ensure robust compensated magnetism, substantial momentum-dependent spin splitting, and energetic 
proximity to the magnetic ground state. 
A material is retained as an altermagnetic candidate if, for at least one 
value of the Hubbard parameter $U$, it simultaneously satisfies the 
following conditions: the absolute magnetization ($M_{abs}$) exceeds $0.1\,\mu_B$/cell, 
confirming the presence of magnetic order; the net magnetization ($M_{tot}$) remains 
below $0.01\,\mu_B$/cell, ensuring full spin compensation and excluding 
ferromagnetic and ferrimagnetic states; the maximum spin splitting 
$\Delta E^{\uparrow\downarrow}_{max}$ exceeds $5$~meV, reflecting a 
sizable momentum-dependent spin splitting characteristic of altermagnetism; 
the spin splitting at the $\Gamma$ point satisfies 
$\Delta E^{\uparrow\downarrow}_{\Gamma,max} <$ 5 meV, which distinguishes altermagnets from compensated ferrimagnets whose spin 
splitting is momentum-independent; and 
finally, the last criterion i.e., $|E_{AM}-E_{GS}|<1$ meV/atom ensures that the altermagnetic state is energetically competitive with the magnetic ground state. Here, $E_{AM}$ denotes the total energy of an altermagnetic configuration that satisfies the first four criteria, while $E_{GS}$ is the total energy of the lowest-energy magnetic configuration among all configurations considered for the same value of $U$.

For the screening, we develop an automated workflow within the AiiDA framework~\cite{huber2020aiida, uhrin2021workflows}, built upon the existing workflows for DFT calculations with \textsc{Quantum ESPRESSO}~\cite{giannozzi2009quantum, giannozzi2017advanced,qe_exa_2020, huber2020aiida,uhrin2021workflows}. The overall procedure is summarized in Fig.~\ref{workflow} and consists of three modular components.

\begin{enumerate}[label=(\alph*)]
\item \texttt{AMAnalysisWorkChain}:  given a set of input structure ($\{x_i\}$), the workflow generates all possible spin configurations ($\{x_{s_i}\}$), and, identifies those consistent with altermagnetic symmetry constraints. The workflow was developed by adapting and extending the \texttt{amchecker} code~\cite{10.21468/SciPostPhysCodeb.30,10.21468/SciPostPhysCodeb.30-r1.0}, a symmetry-based Python package designed for detecting altermagnetism in 3D crystalline systems. \textcolor{black}{To adapt the framework for high-throughput screening of 2D 
materials, we implemented automated generation of compensated magnetic configurations and extended the symmetry analysis to include operations 
relevant to 2D altermagnets, explicitly excluding $C_{2z}$ and $M_z$ as connecting symmetries between opposite-spin sublattices.}

\item \texttt{AMPWScanWorkChain} performs DFT+$U$ across multiple values of $U$ ($U_i$) and starting magnetization ($m_i$) for all the symmetry-allowed altermagnetic configurations obtained from \texttt{AMAnalysisWorkChain} and a ferromagnetic configuration ($\{x^{U_i}_{s_{j},m_{k}}\}$). Electronic-structure calculations are performed using the AiiDA-based \texttt{moderate} protocol by Nascimento et al.~\cite{de2026accurate} (more details in Methods). Because magnetic DFT+$U$ calculations can be prone to convergence difficulties, cases in which the self-consistent field cycle fails to converge are automatically restarted using progressively relaxed convergence thresholds, increased by one order of magnitude at each attempt.

\item \texttt{Analysis of GS-AM}: here the results across all $U$ values are aggregated and ground-state altermagnets are finally identified by applying the criteria defined above.

\end{enumerate}
\vspace{0.5em}

Using this workflow, 42 materials are found to stabilize an altermagnetic ground state for at least one value of $U$. A detailed compilation of these materials, including the relevant $U$ values and the corresponding maximum spin splittings, is provided in the Supporting Information. These compounds are then taken forward for more accurate calculations.

\subsection{Low-throughput refinement}

After identifying the ground-state altermagnets through the high-throughput screening protocol, we perform DFT+$U$ electronic structure calculations in which the Hubbard-$U$ parameters are determined self-consistently within DFPT~\cite{baroni1987green,timrov2022hp} using a dedicated AiiDA workflow~\cite{bastonero2025first} (see Methods for details). Of the 42 candidate altermagnets identified at the scanning stage, self-consistent $U$ values are successfully calculated for 36 materials; the remaining 6 are excluded due to computational or numerical issues discussed in the Methods section. For these 36 materials, we subsequently carry out DFT+$U$ calculations, with the self-consistent $U$, within our automated workflow to assess the stability of the altermagnetic ground states in them. For systems retaining altermagnetic order at this stage, full structural relaxations are performed and the magnetic properties re-evaluated. During this refinement procedure, three materials are excluded. Specifically, Co$_2$NO$_6$ and Co$_2$TaTe$_2$ converge to ferrimagnetic and ferromagnetic ground states, respectively, both exhibiting finite net magnetization. The case of CoH$_2$SeO$_4$ is particularly interesting. The structural relaxation destroys the symmetry connecting the opposite spin sublattice. Therefore, although the net magnetization is compensated, the material lacks the the characteristic symmetry-protected spin degenerate nodes of conventional altermagnets. The resulting spin-split bands more closely resemble those of a compensated ferrimagnet
~\cite{mazin_editorial_2022,yuan_nonrelativistictic_2023,spaldin2026there}. Following this refinement process, 24 materials are confirmed to possess stable altermagnetic ground states. Thus, of the 42 ground-state altermagnets initially identified for at least one value within the scanned Hubbard-$U$ range, only 57\% retain altermagnetic order when self-consistently determined $U$ values are employed\textemdash highlighting the importance of an accurate 
Hubbard-$U$ determination for reliable prediction of altermagnetic phases. We refer to these 24 confirmed systems as refined ground-state altermagnets.

\begin{figure}
\centerline{\includegraphics[scale=0.475]{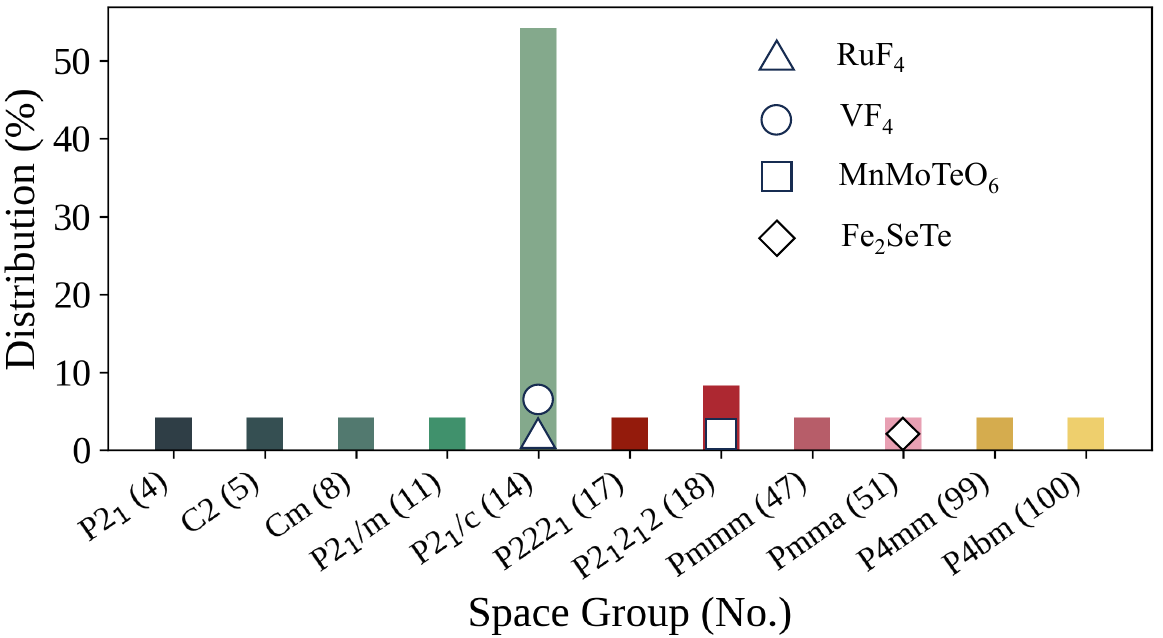}}
\caption{\textbf{Space group distribution of 2D altermagnets.} Distribution of  ground state 2D altermagnetic materials across different space group is shown, with different color of the bar chart denoting different space groups. Previously reported altermagnets\textemdash RuF$_4$~\cite{sodequist2024two, zeng2024description}, VF$_4$~\cite{sodequist2024two}, MnMoTeO$_6$~\cite{zeng2024description} and Fe$_2$SeTe~\cite{gonzalez2026coexistence}\textemdash are highlighted with triangle, circle, square and rhombus, respectively.}
\label{distribution}
\end{figure}

\begin{figure}
\centerline{\includegraphics[scale=0.6]{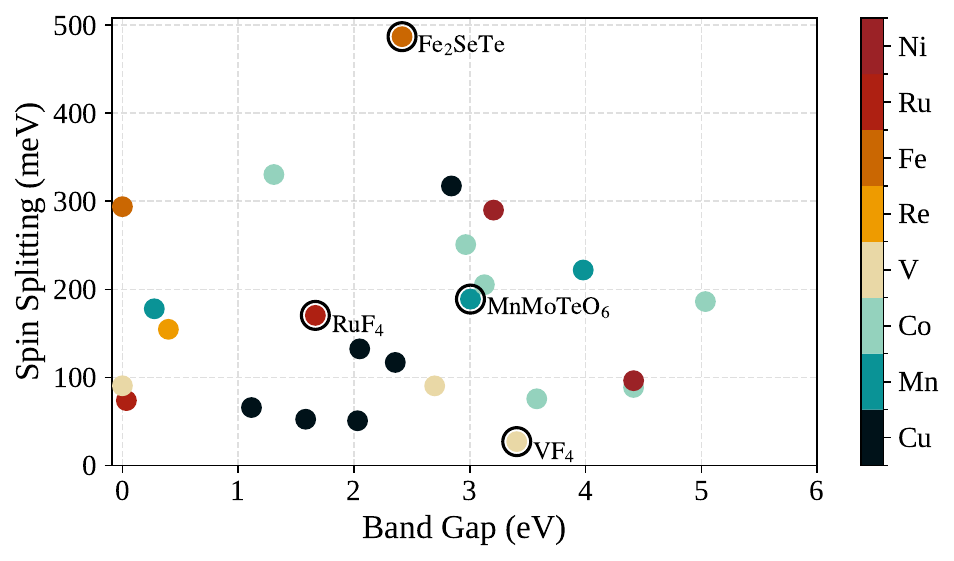}}
\caption{\textbf{Maximum spin splitting in refined ground-state altermagnets.} Scatter plot of the 24 refined ground-state altermagnets identified in this work, showing their maximum spin splitting between the spin-up and spin-down bands (vertical axis) and electronic band gaps (horizontal axis). Each colored circle represents a distinct material, with the color indicating the magnetic element it contains, as specified by the color bar. Four previously reported 2D altermagnets\textemdash RuF$_4$~\cite{sodequist2024two, zeng2024description}, VF$_4$~\cite{sodequist2024two}, MnMoTeO$_6$\cite{zeng2024description}, and Fe$_2$SeTe~\cite{gonzalez2026coexistence} are highlighted with black outlines, while the remaining points correspond to newly identified candidates.}
\label{splitting_gap}
\end{figure}

\begin{figure*}
\centerline{\includegraphics[scale=0.575]{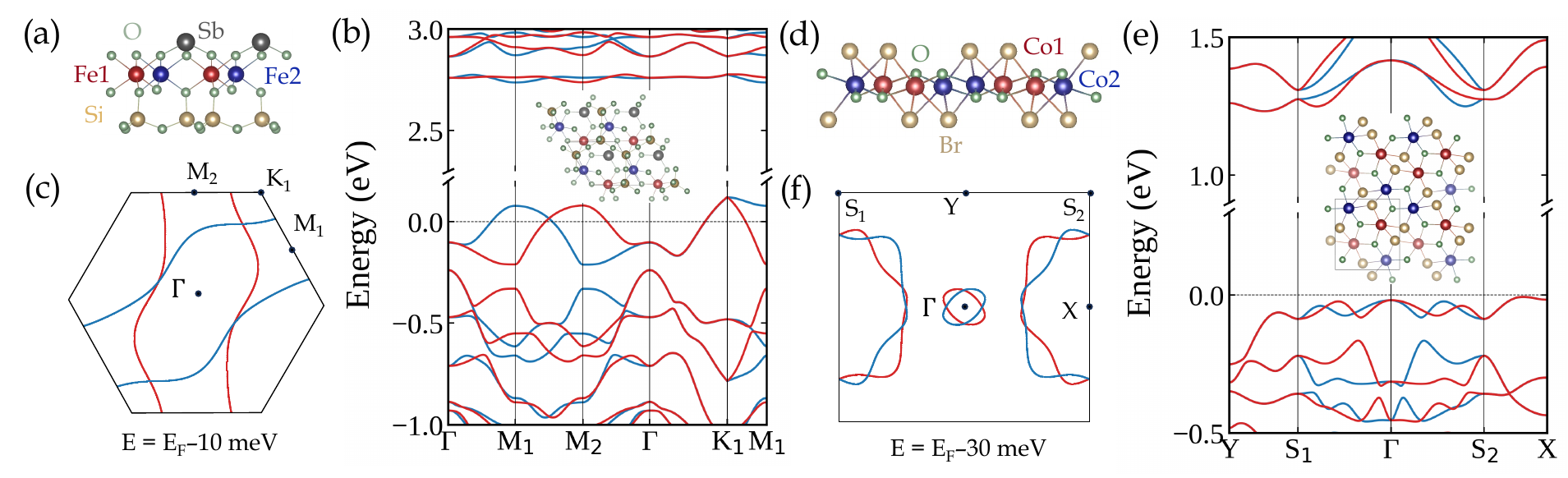}}
\caption{\textbf{Electronic structure for selected 2D altermagnets.} (a) Side view of crystal structure of Fe$_2$Si$_2$SbO$_9$. (b) Spin polarized electronic band structure revealing the characteristic altermagnetic spin splitting; the inset displays the corresponding top view of the crystal structure. (c) Constant energy contour evaluated 10meV below the Fermi level. Panels (d)–(f) present crystal structure, band structure and constant-energy contour computed 30 meV below the Fermi level in case of CoBrO. Fermi level is aligned to the valence-band maximum. Red and blue colors denote contributions from up and down spin channels, respectively.}
\label{bands}
\end{figure*}


For these refined ground-state altermagnets, we further compute their electronic band structures, together with key properties including the maximum spin splitting. In Table~\ref{Table_details}, we summarize the properties of the identified 24 ground-state altermagnetic candidates. The table includes the chemical formula, space group, self-consistent Hubbard-$U$ parameters for the relevant transition-metal $d$-orbitals, the electronic band gap, and spin splitting obtained from DFT+$U$ calculations using the corresponding self-consistent $U$ values. Of these 24 2D altermagnets, only two materials exhibit metallic or semimetallic behavior characterized by electronic states at the Fermi level arising from band crossings or band touching, while the remaining materials are gapped. For insulators, the Fermi level is referenced to the valence-band maximum. The quantity ``maximum spin splitting" is defined as the largest energy separation between spin-split bands across the entire 2D Brillouin zone, considering all electronic states included in the calculation irrespective of their occupation. In contrast, the ``maximum spin splitting near E$_F$" refers to the largest splitting associated with the highest occupied (partially occupied) electronic states of the insulator (metal). Since the reported maximum spin splitting is evaluated over the full Brillouin zone, it is not restricted to high-symmetry paths. We also report the range of $U$ values for which each material stabilizes an altermagnetic ground state, providing a measure of the robustness of the altermagnetic phase. In addition, we list the energetic stability of the altermagnetic ground state, defined as the energy difference between the altermagnetic configuration and the lowest-energy competing magnetic state, evaluated using the corresponding unrelaxed crystal structures. Negative values of the energetic stability for CuH$_8$N$_2$(OF$_2$)$_2$, SrCo$_2$Te$_3$(ClO$_4$)$_2$, V$_3$(H$_3$O$_5$)$_2$, and Al$_2$CuCl$_8$ indicate that the altermagnetic state for these materials is not the lowest-energy state. However, because their energy lies within 1 meV/atom of the lowest-energy state, these materials are also classified as refined ground-state altermagnets according to the criterion adopted in our workflow.\\

Among the proposed materials we identify four materials---RuF$_4$~\cite{sodequist2024two}, VF$_4$~\cite{sodequist2024two}, MnMoTeO$_6$~\cite{zeng2024description} and Fe$_2$SeTe~\cite{gonzalez2026coexistence}\textemdash that had already been proposed as 2D altermagnets. We note that while Fe$_2$SeTe is a substitutionally disordered alloy, here we consider a perfect crystal. 
Nevertheless, local structural probe experiments reveal that Se and Te occupy distinct heights and form inequivalent Fe–chalcogen bond geometries~\cite{louca2010local}, indicating that the local structure might be reasonably approximated by an ordered crystal. The calculated electronic structure of MnMoTeO$_6$ is in qualitative agreement with previous work~\cite{zeng2024description}. In contrast, RuF$_4$~\cite{sodequist2024two}, VF$_4$~\cite{sodequist2024two} and Fe$_2$SeTe~\cite{gonzalez2026coexistence} exhibit noticeable differences relative to earlier DFT+$U$ studies. We attribute these discrepancies primarily to differences in the relaxed structural parameters and the adopted Hubbard-$U$ values. Specifically, the optimized in-plane lattice constants are found to deviate from the reported values by approximately 3.4\%, 4\%, and 17\% for RuF$_4$, VF$_4$, and Fe$_2$SeTe, respectively. We further note that several previously proposed 2D altermagnets~\cite{sodequist2024two,zeng2024description,zeng2024bilayer,wang2025pentagonal} are not recovered in our screening, either because they are not included in the MC2D database or because they were excluded at an earlier stage of our selection pipeline. For instance, AgF$_2$ ~\cite{sodequist2024two} is removed due to constraints on the allowed transition-metal species.

In Figure~\ref{distribution}, we demonstrate the distribution of the refined ground-state altermagnets across different space groups, indicated by different colors. Space-group assignments are determined using \texttt{spglib}~\cite{spglib, spglibv2} with a tolerance of 0.01. A significant fraction of the identified compounds crystallize in the monoclinic P2$_1$/c space group. Previously reported 2D altermagnets are additionally marked using distinct symbols. Figure~\ref{splitting_gap} highlights the diversity of the refined ground-state altermagnets identified in this work in terms of their electronic band gaps and spin splittings. The maximum splitting between spin-up and spin-down bands is shown for all 24 compounds as a function of band gap. The color coding reflects the magnetic element contained in each material, as specified by the colorbar. Previously explored 2D altermagnets are marked by black outlines, while the remaining materials constitute newly identified candidates.

Although our screening identifies several previously unexplored 2D altermagnets with sizable spin splittings (Table~\ref{Table_details}), we briefly highlight two representative material candidates with interesting electronic dispersions, namely Fe$_2$Si$_2$SbO$_9$ and CoBrO. Complete electronic structure data and additional results like spin space group~\cite{yu2026identifying}, symmetry operations connecting opposite spin sublattices for all identified materials are provided in the Supporting Information. Figure~\ref{bands} presents the crystal structures, spin-polarized electronic band structures, and Fermi surfaces of Fe$_2$Si$_2$SbO$_9$ [panels (a)–(c)] and CoBrO [panels (d)–(f)]. The side view of the monolayers for these two materials are depicted in panels (a) and (d), respectively.  On the other hand the top views are shown as insets in (b) and (d). The red and blue colors in band structure and Fermi surface plots denote contributions from the spin-up and spin-down channels, respectively. Fe$_2$Si$_2$SbO$_9$ exhibits metallic behavior, as can be seen from its electronic dispersion in Figure~\ref{bands} (b), with a splitting of 292 meV near the Fermi level. On the other hand, CoBrO is insulating with a band gap of approximately 1.31 eV [Figure~\ref{bands} (e)] and exhibits a spin splitting of 66 meV near E$_F$. Fermi surfaces [(c) and (f)] in both of these compounds demonstrate that the spin splitting changes sign across two nodal planes in momentum space, characteristic of $d$-wave altermagnetic order.

\section{Summary and outlook}

In this work, we carried out a systematic high-throughput search for 2D ground-state altermagnets by combining symmetry-guided screening with 
high-precision first-principles calculations. A key methodological distinction of our approach is the inversion of the conventional workflow: 
rather than determining magnetic ground states first and classifying them afterward, we perform symmetry analysis across all materials in the MC2D 
database before any ground-state calculations, ensuring that no symmetry-permitted altermagnets are missed due to computational constraints. Starting from 2,710 optimized layered compounds, successive filtering based on chemical composition, symmetry compatibility with 
altermagnetism, and exfoliation feasibility reduced the search space to 79 promising candidates. Comprehensive DFT+$U$ calculations over a range of Hubbard-$U$ values and initial magnetic configurations identified 42 
materials exhibiting altermagnetic ground states for at least one value of $U$. Following self-consistent determination of the Hubbard-$U$ parameters using DFPT and re-evaluation of the magnetic and electronic ground states for the relaxed structures, we arrive at a final set of 24 robust 2D altermagnetic candidates\textemdash corresponding to $\sim$0.89\% of the materials in the MC2D database, a significantly larger fraction than identified by prior searches on the same database~\cite{haddadi2026exploring}.

For these materials, we performed detailed electronic-structure analyses, including band dispersions and momentum-dependent spin splitting, and quantified the maximum spin splitting across the full Brillouin zone. Four of the identified compounds have been previously reported as candidate 2D altermagnets, while the remaining 20 substantially expand the current pool of experimentally and theoretically relevant systems. The chemical and structural diversity of the identified altermagnets opens several exciting 
directions for future investigation. Among the identified systems, ReCl$_3$ stands out for its 
small bandwidth near the Fermi level, making it a promising platform to study the interplay between altermagnetism and strong electronic correlations\textemdash a largely unexplored regime in the context of 2D magnetism. Equally noteworthy is the identification of Fe$_2$SeTe as a 2D altermagnet, a compound that can be viewed as a substitutionally doped analogue of FeSe\textemdash a celebrated Fe-based superconductor~\cite{he2013phase,lee2014interfacial}. This raises the intriguing possibility that altermagnetism and 
superconductivity may coexist or compete in this family, with Fe$_2$SeTe serving as a tunable bridge between the two phases. More broadly, the 
exfoliable nature of our candidates makes them immediately relevant for van der Waals heterostructure engineering, where proximity effects, gate-voltage control, and moiré engineering can be used to further tune the altermagnetic order and spin splitting.

Beyond the specific materials identified here, this work establishes a fully automated AiiDA-based workflow that, starting from the crystal structure, enables identification of ground-state altermagnets and evaluation of spin-splitting characteristics based on DFT+$U$ calculations. The workflow is scalable and transferable, and can be readily extended to other classes of magnetic quantum materials. Taken together, our results significantly enlarge the landscape of candidate 2D altermagnets and provide a practical 
and reproducible foundation for future experimental realization and device-oriented investigations.\\

\section*{Methods}
The DFT calculations are carried out using
\textsc{Quantum ESPRESSO}~\cite{giannozzi2009quantum, giannozzi2017advanced,qe_exa_2020} employing the Perdew--Burke--Ernzerhof (PBE) exchange--correlation functional~\cite{perdew1996generalized} and the Standard Solid-State Pseudopotentials (SSSP) library~\cite{vanderbilt1990soft,Blochl1994projector,Goedecker1996separable,Hamann2013optimized,willand2013norm,kucukbenli2014projector,dal2014pseudopotentials,garrity2014pseudopotentials,schlipf2015optimization,lejaeghere2016reproducibility,prandini2018precision}. Brillouin-zone sampling was generated using the ``moderate'' protocol of the AiiDA PW WorkChain~\cite{de2026accurate}, which employs the smallest $k$-point mesh consistent with a reciprocal-space sampling density of $0.15\,\mathrm{\AA}^{-1}$. Pseudopotentials were selected from the SSSP library, and the wavefunction and charge-density cutoffs were chosen according to the SSSP recommendations, taking the maximum recommended value among the constituent elements of each structure. A Coulomb cutoff is used to avoid spurious interactions between periodic replicas and thus simulate the correct boundary conditions for 2D systems~\cite{sohier_prb_2017}.

The high-throughput DFT+$U$ calculations performed over the scanned range of Hubbard-$U$ values employed Methfessel-Paxton~\cite{methfessel1989high} smearing to ensure robust convergence across the diverse set of candidate materials. For the 
subsequent calculations, in which the Hubbard-$U$ parameters are determined self-consistently and the magnetic and electronic ground states are re-evaluated, we adopted Marzari-Vanderbilt cold smearing~\cite{marzari1999thermal}, 
which provides a more accurate description of the electronic structure and is better suited for the precise determination of ground-state properties.

Hubbard-$U$ parameters were computed using DFPT~\cite{baroni1987green,gonze1997first,timrov2022hp} as implemented in \textsc{Quantum ESPRESSO}. To obtain self-consistent $U$ values for 42 materials, we employed the automated AiiDA Hubbard workflow described in Ref.~\cite{bastonero2025first}, using the balanced protocol. The workflow successfully converged for 24 materials, yielding self-consistent Hubbard-$U$ parameters. For the remaining 18 systems, where the automated workflow did not reach convergence, the self-consistent determination of $U$ was carried out manually using the same DFPT methodology. In these cases, successive DFPT calculations were iterated until the change in the Hubbard-$U$ value between consecutive iterations was less than 0.1 eV. We note that self-consistent Hubbard-$U$ calculations could not be successfully completed for a small subset of materials, despite their identification as potential altermagnetic candidates within our workflow. For example, TlCr$_{4}$BiO$_{14}$
 was excluded from the refined calculations due to the prohibitively high computational cost associated with its large unit cell. For FePSe$_3$, the self-consistent determination of the Hubbard-$U$ parameter for the Fe-3$d$ states exhibited significant fluctuations and did not converge reliably. In the case of Cu$_2$Fe$_4$S$_7$, the self-consistently computed Hubbard-$U$ parameter for Cu was found to be unphysical. This behaviour can be attributed to the closed $3d$ shell electronic configuration of Cu$^{+1}$, for which the response matrix becomes ill-conditioned and numerically unstable within the linear-response formalism~\cite{yu2014communication}. A similar behaviour was also observed for Ni-3$d$, Fe-3$d$ and Co-3$d$ electrons in V$_2$Ni(PO$_5$)$_2$, Ba(Fe$_4$As)$_2$ and TaCoTe$_2$, respectively.\\


\section*{Data Availability}
The data that support the findings of this study will be made openly available on the Materials Cloud~\cite{MC_dataset} at the time of publication. The source code of AiiDA-Altermagnets will be made publicly available on GitHub with MIT license at \href{https://github.com/ARGO-SISSA/AiiDA-altermagnets}{github.com/ARGO-SISSA/AiiDA-altermagnets}.

\section*{Acknowledgement}

A.B. thanks Md.\ Afsar Reja and Nayana Devaraj for useful discussions. This study is  funded by the European Union–NextGenerationEU, through the PRIN Project “Simultaneous electrical control of spin and valley polarization in van der Waals magnetic materials” (SECSY–CUP Grant No. J53D23001400001, PNRR
Investimento M4.C2.1.1). N.M. and A.M. acknowledge partial support from the European Commission through the Centre of Excellence ``MaX - Materials Design at the Exascale'' (HORIZON-EUROHPC, Grant No. 101093374).  The views and opinions expressed are solely those of the authors and do not necessarily reflect those of the European Union, nor can the European Union be held responsible for them.

\bibliography{references}

@article{sohier_prb_2017,
  title = {Density functional perturbation theory for gated two-dimensional heterostructures: Theoretical developments and application to flexural phonons in graphene},
  author = {Sohier, Thibault and Calandra, Matteo and Mauri, Francesco},
  journal = {Phys. Rev. B},
  volume = {96},
  issue = {7},
  pages = {075448},
  numpages = {21},
  year = {2017},
  month = {Aug},
  publisher = {American Physical Society},
  doi = {10.1103/PhysRevB.96.075448},

}

@Misc{MC_dataset,
  author    = {Bose, Anumita and Manko, Nataliia and Gibertini, Marco and Marrazzo, Antimo},
  title     = {{2D} altermagnets database},
  year      = {2026},
  doi       = {XXX},
  publisher = {Materials Cloud},
}

@article{vsmejkal2022emerging,
  title={Emerging research landscape of altermagnetism},
  author={{\v{S}}mejkal, Libor and Sinova, Jairo and Jungwirth, Tomas},
  journal={Physical Review X},
  volume={12},
  number={4},
  pages={040501},
  year={2022},
  publisher={APS},
  url={https://doi.org/10.1103/PhysRevX.12.040501}
}

@article{vsmejkal2022beyond,
  title={Beyond conventional ferromagnetism and antiferromagnetism: A phase with nonrelativistic spin and crystal rotation symmetry},
  author={{\v{S}}mejkal, Libor and Sinova, Jairo and Jungwirth, Tomas},
  journal={Physical Review X},
  volume={12},
  number={3},
  pages={031042},
  year={2022},
  publisher={APS},
  url={https://doi.org/10.1103/PhysRevX.12.031042}
}

@article{bai2024altermagnetism,
  title={Altermagnetism: Exploring new frontiers in magnetism and spintronics},
  author={Bai, Ling and Feng, Wanxiang and Liu, Siyuan and {\v{S}}mejkal, Libor and Mokrousov, Yuriy and Yao, Yugui},
  journal={Advanced Functional Materials},
  volume={34},
  number={49},
  pages={2409327},
  year={2024},
  publisher={Wiley Online Library},
  url={https://doi.org/10.1002/adfm.202409327}
}

@article{krempasky2024altermagnetic,
  title={Altermagnetic lifting of Kramers spin degeneracy},
  author={Krempask{\`y}, Juraj and {\v{S}}mejkal, L and D’souza, SW and Hajlaoui, M and Springholz, G and Uhl{\'\i}{\v{r}}ov{\'a}, K and Alarab, F and Constantinou, PC and Strocov, V and Usanov, D and others},
  journal={Nature},
  volume={626},
  number={7999},
  pages={517--522},
  year={2024},
  publisher={Nature Publishing Group UK London},
  url={https://doi.org/10.1038/s41586-023-06907-7}
}

@article{fender2025altermagnetism,
  title={Altermagnetism: A chemical perspective},
  author={Fender, Shannon S and Gonzalez, Oscar and Bediako, D Kwabena},
  journal={Journal of the American Chemical Society},
  volume={147},
  number={3},
  pages={2257--2274},
  year={2025},
  publisher={ACS Publications},
  url={https://doi.org/10.1021/jacs.4c14503}
}

@article{tamang2025altermagnetism,
  title={Altermagnetism and altermagnets: A brief review},
  author={Tamang, Rupam and Gurung, Shivraj and Rai, Dibya Prakash and Brahimi, Samy and Lounis, Samir},
  journal={Magnetism},
  volume={5},
  number={3},
  pages={17},
  year={2025},
  publisher={MDPI},
  url={https://doi.org/10.1063/5.0198285}
}

@article{zhang2025theory,
  title={Theory of anisotropic magnetoresistance in altermagnets and its applications},
  author={Zhang, Xian-Peng and Feng, Wanxiang and Zhang, Run-Wu and Fan, Xiaolong and Wang, Xiangrong and Yao, Yugui},
  journal={Physical Review Letters},
  volume={135},
  number={26},
  pages={266706},
  year={2025},
  publisher={APS},
  url={https://doi.org/10.1103/32zc-ggjy}
}

@article{gonzalez2024anisotropic,
  title={Anisotropic magnetoresistance in altermagnetic MnTe},
  author={Gonzalez Betancourt, Ruben Dario and Zubáč, Jan and Geishendorf, Kevin and Ritzinger, Philipp and Růžičková, Barbora and Kotte, Tommy and Železný, Jakub and Olejník, Kamil and Springholz, Gunther and Büchner, Bernd and Thomas, Andy and Výborný, Karel and Jungwirth, Tomas and Reichlová, Helena and Kriegner, Dominik},
  journal={npj Spintronics},
  volume={2},
  number={1},
  pages={45},
  year={2024},
  publisher={Nature Publishing Group UK London},
  url={https://doi.org/10.1038/s44306-024-00046-z}
}

@article{gao2025ai,
  title={AI-accelerated discovery of altermagnetic materials},
  author={Gao, Ze-Feng and Qu, Shuai and Zeng, Bocheng and Liu, Yang and Wen, Ji-Rong and Sun, Hao and Guo, Peng-Jie and Lu, Zhong-Yi},
  journal={National Science Review},
  url={https://doi.org/10.1093/nsr/nwaf066},
  volume={12},
  number={4},
  pages={nwaf066},
  year={2025},
  publisher={Oxford University Press}
}

@article{song2025altermagnets,
  title={Altermagnets as a new class of functional materials},
  author={Song, Cheng and Bai, Hua and Zhou, Zhiyuan and Han, Lei and Reichlova, Helena and Dil, J Hugo and Liu, Junwei and Chen, Xianzhe and Pan, Feng},
  journal={Nature Reviews Materials},
  url={https://doi.org/10.1038/s41578-025-00779-1},
  volume={10},
  number={6},
  pages={473--485},
  year={2025},
  publisher={Nature Publishing Group UK London}
}

@article{guo2025spin,
  title={Spin-Polarized Antiferromagnets for Spintronics},
  author={Guo, Zhenzhou and Wang, Xiaotian and Wang, Wenhong and Zhang, Gang and Zhou, Xiaodong and Cheng, Zhenxiang},
  journal={Advanced Materials},
  volume={37},
  number={36},
  pages={2505779},
  year={2025},
  publisher={Wiley Online Library},
  url={https://doi.org/10.1002/adma.202505779}
}

@article{chen2024emerging,
  title={Emerging antiferromagnets for spintronics},
  author={Chen, Hongyu and Liu, Li and Zhou, Xiaorong and Meng, Ziang and Wang, Xiaoning and Duan, Zhiyuan and Zhao, Guojian and Yan, Han and Qin, Peixin and Liu, Zhiqi},
  journal={Advanced Materials},
  volume={36},
  number={14},
  pages={2310379},
  year={2024},
  publisher={Wiley Online Library},
  url={https://doi.org/10.1002/adma.202310379}
}

@article{vsmejkal2022giant,
  title={Giant and tunneling magnetoresistance in unconventional collinear antiferromagnets with nonrelativistic spin-momentum coupling},
  author={{\v{S}}mejkal, Libor and Hellenes, Anna Birk and Gonz{\'a}lez-Hern{\'a}ndez, Rafael and Sinova, Jairo and Jungwirth, Tomas},
  journal={Physical Review X},
  volume={12},
  number={1},
  pages={011028},
  year={2022},
  publisher={APS},
  url={https://doi.org/10.1103/PhysRevX.12.011028}
}

@article{gonzalez2021efficient,
  title={Efficient electrical spin splitter based on nonrelativistic collinear antiferromagnetism},
  author={Gonz{\'a}lez-Hern{\'a}ndez, Rafael and {\v{S}}mejkal, Libor and V{\`y}born{\`y}, Karel and Yahagi, Yuta and Sinova, Jairo and Jungwirth, Tom{\'a}{\v{s}} and {\v{Z}}elezn{\`y}, Jakub},
  journal={Physical Review Letters},
  volume={126},
  number={12},
  pages={127701},
  year={2021},
  publisher={APS},
  url={https://doi.org/10.1103/PhysRevLett.126.127701}
}

@article{karube2022observation,
  title={Observation of spin-splitter torque in collinear antiferromagnetic {RuO$_2$}},
  author={Karube, Shutaro and Tanaka, Takahiro and Sugawara, Daichi and Kadoguchi, Naohiro and Kohda, Makoto and Nitta, Junsaku},
  journal={Physical Review Letters},
  volume={129},
  number={13},
  pages={137201},
  year={2022},
  publisher={APS},
  url={https://doi.org/10.1103/PhysRevLett.129.137201}
}

@article{bai2022observation,
  title={Observation of spin splitting torque in a collinear antiferromagnet {RuO$_2$}},
  author={Bai, Hua and Han, Lei and Feng, XY and Zhou, YJ and Su, RX and Wang, Qian and Liao, LY and Zhu, WX and Chen, XZ and Pan, Feng and others},
  journal={Physical Review Letters},
  volume={128},
  number={19},
  pages={197202},
  year={2022},
  publisher={APS},
  url={https://doi.org/10.1103/PhysRevLett.128.197202}
}

@article{bhattarai2025high,
  title={High-throughput screening of altermagnetic materials},
  author={Bhattarai, Romakanta and Minch, Peter and Rhone, Trevor David},
  journal={Physical Review Materials},
  volume={9},
  number={6},
  pages={064403},
  year={2025},
  publisher={APS},
  url={https://doi.org/10.1103/PhysRevMaterials.9.064403}
}

@article{amin2024nanoscale,
  title={Nanoscale imaging and control of altermagnetism in {MnTe}},
  author={Amin, OJ and Dal Din, A and Golias, E and Niu, Y and Zakharov, A and Fromage, SC and Fields, CJB and Heywood, SL and Cousins, RB and Maccherozzi, F and others},
  journal={Nature},
  volume={636},
  number={8042},
  pages={348--353},
  year={2024},
  publisher={Nature Publishing Group UK London},
  url={https://doi.org/10.1038/s41586-024-08234-x}
}

@article{fedchenko2024observation,
  title={Observation of time-reversal symmetry breaking in the band structure of altermagnetic {RuO$_2$}},
  author={Fedchenko, Olena and Min{\'a}r, Jan and Akashdeep, Akashdeep and D’Souza, Sunil Wilfred and Vasilyev, Dmitry and Tkach, Olena and Odenbreit, Lukas and Nguyen, Quynh and Kutnyakhov, Dmytro and Wind, Nils and others},
  journal={Science Advances},
  volume={10},
  number={5},
  pages={eadj4883},
  year={2024},
  publisher={American Association for the Advancement of Science},
  url={https://doi.org/10.1126/sciadv.adj4883}
}

@article{guo2024direct,
  title={Direct and inverse spin splitting effects in altermagnetic {RuO$_2$}},
  author={Guo, Yaqin and Zhang, Jing and Zhu, Zengtai and Jiang, Yuan-yuan and Jiang, Longxing and Wu, Chuangwen and Dong, Jing and Xu, Xing and He, Wenqing and He, Bin and others},
  journal={Advanced Science},
  volume={11},
  number={25},
  pages={2400967},
  year={2024},
  publisher={Wiley Online Library},
  url={https://doi.org/10.1002/advs.202400967}
}

@article{reimers2024direct,
  title={Direct observation of altermagnetic band splitting in {CrSb} thin films},
  author={Reimers, Sonka and Odenbreit, Lukas and {\v{S}}mejkal, Libor and Strocov, Vladimir N and Constantinou, Procopios and Hellenes, Anna B and Jaeschke Ubiergo, Rodrigo and Campos, Warlley H and Bharadwaj, Venkata K and Chakraborty, Atasi and others},
  journal={Nature Communications},
  volume={15},
  number={1},
  pages={2116},
  year={2024},
  publisher={Nature Publishing Group UK London},
  url={https://doi.org/10.1038/s41467-024-46476-5}
}

@article{ding2024large,
  title={Large band splitting in $g$-wave altermagnet {CrSb}},
  author={Ding, Jianyang and Jiang, Zhicheng and Chen, Xiuhua and Tao, Zicheng and Liu, Zhengtai and Li, Tongrui and Liu, Jishan and Sun, Jianping and Cheng, Jinguang and Liu, Jiayu and others},
  journal={Physical Review Letters},
  volume={133},
  number={20},
  pages={206401},
  year={2024},
  publisher={APS},
  url={https://doi.org/10.1103/PhysRevLett.133.206401}
}

@article{lee2016ising,
  title={Ising-type magnetic ordering in atomically thin {FePS$_3$}},
  author={Lee, Jae-Ung and Lee, Sungmin and Ryoo, Ji Hoon and Kang, Soonmin and Kim, Tae Yun and Kim, Pilkwang and Park, Cheol-Hwan and Park, Je-Geun and Cheong, Hyeonsik},
  journal={Nano Letters},
  volume={16},
  number={12},
  pages={7433--7438},
  year={2016},
  publisher={ACS Publications},
  url={http://dx.doi.org/10.1021/acs.nanolett.6b03052}
}

@article{gong2017discovery,
  title={Discovery of intrinsic ferromagnetism in two-dimensional van der Waals crystals},
  author={Gong, Cheng and Li, Lin and Li, Zhenglu and Ji, Huiwen and Stern, Alex and Xia, Yang and Cao, Ting and Bao, Wei and Wang, Chenzhe and Wang, Yuan and others},
  journal={Nature},
  volume={546},
  number={7657},
  pages={265--269},
  year={2017},
  publisher={Nature Publishing Group UK London},
  url={https://doi.org/10.1038/nature22060}
}

@article{huang2017layer,
  title={Layer-dependent ferromagnetism in a van der Waals crystal down to the monolayer limit},
  author={Huang, Bevin and Clark, Genevieve and Navarro-Moratalla, Efr{\'e}n and Klein, Dahlia R and Cheng, Ran and Seyler, Kyle L and Zhong, Ding and Schmidgall, Emma and McGuire, Michael A and Cobden, David H and others},
  journal={Nature},
  volume={546},
  number={7657},
  pages={270--273},
  year={2017},
  publisher={Nature Publishing Group UK London},
  url={https://doi.org/10.1038/nature22391}
}

@article{burch2018magnetism,
  title={Magnetism in two-dimensional van der Waals materials},
  author={Burch, Kenneth S and Mandrus, David and Park, Je-Geun},
  journal={Nature},
  volume={563},
  number={7729},
  pages={47--52},
  year={2018},
  publisher={Nature Publishing Group UK London},
  url={https://doi.org/10.1038/s41586-018-0631-z}
}

@article{gong2019two,
  title={Two-dimensional magnetic crystals and emergent heterostructure devices},
  author={Gong, Cheng and Zhang, Xiang},
  journal={Science},
  volume={363},
  number={6428},
  pages={eaav4450},
  year={2019},
  publisher={American Association for the Advancement of Science},
  url={https://doi.org/10.1126/science.aav4450}
}

@article{li2019intrinsic,
  title={Intrinsic van der Waals magnetic materials from bulk to the {2D} limit: new frontiers of spintronics},
  author={Li, Hui and Ruan, Shuangchen and Zeng, Yu-Jia},
  journal={Advanced Materials},
  volume={31},
  number={27},
  pages={1900065},
  year={2019},
  publisher={Wiley Online Library},
  url={https://doi.org/10.1002/adma.201900065}
}

@article{khan2020recent,
  title={Recent breakthroughs in two-dimensional van der Waals magnetic materials and emerging applications},
  author={Khan, Yahya and Obaidulla, Sk Md and Habib, Mohammad Rezwan and Gayen, Anabil and Liang, Tao and Wang, Xuefeng and Xu, Mingsheng},
  journal={Nano Today},
  volume={34},
  pages={100902},
  year={2020},
  publisher={Elsevier},
  url={https://doi.org/10.1016/j.nantod.2020.100902}
}

@article{liu2023emergent,
  title={Emergent, Non-Aging, Extendable, and Rechargeable Exchange Bias in {2D} {Fe$_3$GeTe$_2$} Homostructures Induced by Moderate Pressuring},
  author={Liu, Caixing and Zhang, Huisheng and Zhang, Shunhong and Hou, De and Liu, Yonglai and Wu, Hanqing and Jiang, Zhongzhu and Wang, HuaiXiang and Ma, Zongwei and Luo, Xuan and others},
  journal={Advanced Materials},
  volume={35},
  number={1},
  pages={2203411},
  year={2023},
  publisher={Wiley Online Library},
  url={https://doi.org/10.1002/adma.202203411}
}

@article{novoselov20162d,
  title={{2D} materials and van der Waals heterostructures},
  author={Novoselov, K Sꎬ and Mishchenko, Artem and Carvalho, Alexandra and Castro Neto, AH},
  journal={Science},
  url={https://www.science.org/doi/10.1126/science.aac9439},
  volume={353},
  number={6298},
  pages={aac9439},
  year={2016},
  publisher={American Association for the Advancement of Science}
}

@article{liu2016van,
  title={Van der Waals heterostructures and devices},
  author={Liu, Yuan and Weiss, Nathan O and Duan, Xidong and Cheng, Hung-Chieh and Huang, Yu and Duan, Xiangfeng},
  journal={Nature Reviews Materials},
  url={https://doi.org/10.1038/natrevmats.2016.42},
  volume={1},
  number={9},
  pages={16042},
  year={2016},
  publisher={Nature Publishing Group}
}

@article{chen2019electrically,
  title={Electrically tunable physical properties of two-dimensional materials},
  author={Chen, Xiaolong and Zhou, Zishu and Deng, Bingchen and Wu, Zefei and Xia, Fengnian and Cao, Yi and Zhang, Le and Huang, Wei and Wang, Ning and Wang, Lin},
  journal={Nano Today},
  url={https://doi.org/10.1016/j.nantod.2019.05.005},
  volume={27},
  pages={99--119},
  year={2019},
  publisher={Elsevier}
}

@article{du2021engineering,
  title={Engineering symmetry breaking in {2D} layered materials},
  author={Du, Luojun and Hasan, Tawfique and Castellanos-Gomez, Andres and Liu, Gui-Bin and Yao, Yugui and Lau, Chun Ning and Sun, Zhipei},
  journal={Nature Reviews Physics},
  url={https://doi.org/10.1038/s42254-020-00276-0},
  volume={3},
  number={3},
  pages={193--206},
  year={2021},
  publisher={Nature Publishing Group UK London}
}

@article{nonrelativistic_He2023,
  title = {Nonrelativistic Spin-Momentum Coupling in Antiferromagnetic Twisted Bilayers},
  author = {He, Ran and Wang, Dan and Luo, Nannan and Zeng, Jiang and Chen, Ke-Qiu and Tang, Li-Ming},
  journal = {Phys. Rev. Lett.},
  volume = {130},
  issue = {4},
  pages = {046401},
  numpages = {6},
  year = {2023},
  month = {Jan},
  publisher = {American Physical Society},
  doi = {10.1103/PhysRevLett.130.046401},
  url = {https://link.aps.org/doi/10.1103/PhysRevLett.130.046401}
}

@article{nonrelativistic_Sheoran2024,
  title = {Nonrelativistic spin splittings and altermagnetism in twisted bilayers of centrosymmetric antiferromagnets},
  author = {Sheoran, Sajjan and Bhattacharya, Saswata},
  journal = {Phys. Rev. Mater.},
  volume = {8},
  issue = {5},
  pages = {L051401},
  numpages = {8},
  year = {2024},
  month = {May},
  publisher = {American Physical Society},
  doi = {10.1103/PhysRevMaterials.8.L051401},
  url = {https://link.aps.org/doi/10.1103/PhysRevMaterials.8.L051401}
}

@article{mazin2023induced,
  title={Induced Monolayer Altermagnetism in MnP (S, Se) $ \_3 $ and FeSe},
  author={Mazin, Igor and Gonz{\'a}lez-Hern{\'a}ndez, Rafael and {\v{S}}mejkal, Libor},
  journal={arXiv preprint arXiv:2309.02355},
  year={2023},
  url={https://arxiv.org/abs/2309.02355}
}

@article{yuan_nonrelativistictic_2023,
  title = {Nonrelativistic Spin Splitting at the Brillouin Zone Center in Compensated Magnets},
  author = {Yuan, Lin-Ding and Georgescu, Alexandru B. and Rondinelli, James M.},
  journal = {Phys. Rev. Lett.},
  volume = {133},
  issue = {21},
  pages = {216701},
  numpages = {8},
  year = {2024},
  month = {Nov},
  publisher = {American Physical Society},
  doi = {10.1103/PhysRevLett.133.216701},
  url = {https://link.aps.org/doi/10.1103/PhysRevLett.133.216701}
}

@article{mazin_editorial_2022,
  title = {Editorial: Altermagnetism---A New Punch Line of Fundamental Magnetism},
  author = {Mazin, Igor},
  collaboration = {The PRX Editors},
  journal = {Phys. Rev. X},
  volume = {12},
  issue = {4},
  pages = {040002},
  numpages = {3},
  year = {2022},
  month = {Dec},
  publisher = {American Physical Society},
  doi = {10.1103/PhysRevX.12.040002},
  url = {https://link.aps.org/doi/10.1103/PhysRevX.12.040002}
}

@article{guo2023spin,
  title={Spin-split collinear antiferromagnets: A large-scale ab-initio study},
  author={Guo, Yaqian and Liu, Hui and Janson, Oleg and Fulga, Ion Cosma and van den Brink, Jeroen and Facio, Jorge I},
  journal={Materials Today Physics},
  volume={32},
  pages={100991},
  year={2023},
  publisher={Elsevier},
  url={https://doi.org/10.1016/j.mtphys.2023.100991}
}

@article{zeng2024description,
  title={Description of two-dimensional altermagnetism: Categorization using spin group theory},
  author={Zeng, Sike and Zhao, Yu-Jun},
  journal={Physical Review B},
  volume={110},
  number={5},
  pages={054406},
  year={2024},
  publisher={APS},
  url={https://doi.org/10.1103/PhysRevB.110.054406}
}

@article{sodequist2024two,
  title={Two-dimensional altermagnets from high throughput computational screening: Symmetry requirements, chiral magnons, and spin-orbit effects},
  author={S{\o}dequist, Joachim and Olsen, Thomas},
  journal={Applied Physics Letters},
  volume={124},
  number={18},
  pages={182409},
  year={2024},
  publisher={AIP Publishing},
  url={https://doi.org/10.1063/5.0198285}
}

@article{wang2025pentagonal,
  title={Pentagonal {2D} Altermagnets: Material Screening and Altermagnetic Tunneling Junction Device Application},
  author={Wang, Jianhua and Yang, Xingyue and Yang, Zongmeng and Lu, Jing and Ho, Pin and Wang, Wenhong and Ang, Yee Sin and Cheng, Zhenxiang and Fang, Shibo},
  journal={Advanced Functional Materials},
  pages={2505145},
  year={2025},
  publisher={Wiley Online Library},
  url={https://doi.org/10.1002/adfm.202505145}
}

@article{zeng2024bilayer,
  title={Bilayer stacking A-type altermagnet: A general approach to generating two-dimensional altermagnetism},
  author={Zeng, Sike and Zhao, Yu-Jun},
  journal={Physical Review B},
  volume={110},
  number={17},
  pages={174410},
  year={2024},
  publisher={APS},
  url={https://doi.org/10.1103/PhysRevB.110.174410}
}

@article{sufyan2026high,
  title={High-throughput quantification of altermagnetic band splitting},
  author={Sufyan, Ali and Marfoua, Brahim and Larsson, J Andreas and Van Loon, Erik and Armiento, Rickard},
  journal={Physical Review Materials},
  doi = {10.1103/mmdm-hrj4},
  url = {https://link.aps.org/doi/10.1103/mmdm-hrj4},
  volume={10},
  number={4},
  pages={044407},
  year={2026},
  publisher={APS}
}

@article{haddadi2026exploring,
  title={Exploring the Magnetic Landscape of Easily Exfoliable Two-Dimensional Materials},
  author={Haddadi, Fatemeh and Campi, Davide and Dos Santos, Flaviano Jos{\'e} and Mounet, Nicolas and Ponet, Louis and Marzari, Nicola and Gibertini, Marco},
  journal={ACS Nano},
  volume={20},
  number={18},
  pages={13528--13541},
  year={2026},
  publisher={ACS Publications},
  url={https://doi.org/10.1021/acsnano.5c16067}
}

@article{gonzalez2026coexistence,
  title={Coexistence of $d$-wave altermagnetism and topological states in Janus {FeSeX (X= S, Te)} monolayers},
  author={Gonz{\'a}lez-Garc{\'\i}a, Alvaro and L{\'o}pez-P{\'e}rez, William and Pacheco, Paola and Ram{\'\i}rez-Montes, Luz and Gonz{\'a}lez-Hern{\'a}ndez, Rafael},
  journal={Physical Review Materials},
  doi = {10.1103/3kkj-s5jk},
  url = {https://link.aps.org/doi/10.1103/3kkj-s5jk},
  volume={10},
  number={4},
  pages={044004},
  year={2026},
  publisher={APS}
}

@article{louca2010local,
  title={Local atomic structure of superconducting {FeSe$_{1-x}$Te$_x$}},
  author={Louca, Despina and Horigane, Kazumasa and Llobet, Anna and Arita, Ryotaro and Ji, Sungdae and Katayama, Naoyuki and Konbu, Shun and Nakamura, Kazuma and Koo, T-Y and Tong, Peng and others},
  journal={Physical Review B},
  doi = {10.1103/PhysRevB.81.134524},
  url = {https://link.aps.org/doi/10.1103/PhysRevB.81.134524},
  volume={81},
  number={13},
  pages={134524},
  year={2010},
  publisher={APS}
}

@article{mounet2018two,
  title={Two-dimensional materials from high-throughput computational exfoliation of experimentally known compounds},
  author={Mounet, Nicolas and Gibertini, Marco and Schwaller, Philippe and Campi, Davide and Merkys, Andrius and Marrazzo, Antimo and Sohier, Thibault and Castelli, Ivano Eligio and Cepellotti, Andrea and Pizzi, Giovanni and others},
  journal={Nature Nanotechnology},
  volume={13},
  number={3},
  pages={246--252},
  year={2018},
  publisher={Nature Publishing Group UK London},
  url={https://doi.org/10.1038/s41565-017-0035-5}
}

@article{campi2023expansion,
  title={Expansion of the materials cloud {2D} database},
  author={Campi, Davide and Mounet, Nicolas and Gibertini, Marco and Pizzi, Giovanni and Marzari, Nicola},
  journal={ACS Nano},
  volume={17},
  number={12},
  pages={11268--11278},
  year={2023},
  publisher={ACS Publications},
  url={https://doi.org/10.1021/acsnano.2c11510?urlappend=%3Fref%3DPDF&jav=VoR&rel=cite-as}
}

@article{wang2024electric,
  title={Electric-field-induced switchable two-dimensional altermagnets},
  author={Wang, Dinghui and Wang, Huaiqiang and Liu, Lulu and Zhang, Junting and Zhang, Haijun},
  journal={Nano Letters},
  volume={25},
  number={1},
  pages={498--503},
  year={2024},
  publisher={ACS Publications},
  url={https://doi.org/10.1021/acs.nanolett.4c05384}
}

@article{zhang2024predictable,
  title={Predictable gate-field control of spin in altermagnets with spin-layer coupling},
  author={Zhang, Run-Wu and Cui, Chaoxi and Li, Runze and Duan, Jingyi and Li, Lei and Yu, Zhi-Ming and Yao, Yugui},
  journal={Physical Review Letters},
  volume={133},
  number={5},
  pages={056401},
  year={2024},
  publisher={APS},
  url={https://doi.org/10.1103/PhysRevLett.133.056401}
}

@article{liu2024twisted,
  title={Twisted magnetic van der Waals bilayers: An ideal platform for altermagnetism},
  author={Liu, Yichen and Yu, Junxi and Liu, Cheng-Cheng},
  journal={Physical Review Letters},
  volume={133},
  number={20},
  pages={206702},
  year={2024},
  publisher={APS},
  url={https://doi.org/10.1103/PhysRevLett.133.206702}
}

@article{pan2024general,
  title={General stacking theory for altermagnetism in bilayer systems},
  author={Pan, Baoru and Zhou, Pan and Lyu, Pengbo and Xiao, Huaping and Yang, Xuejuan and Sun, Lizhong},
  journal={Physical Review Letters},
  volume={133},
  number={16},
  pages={166701},
  year={2024},
  publisher={APS},
  url={https://doi.org/10.1103/PhysRevLett.133.166701}
}

@article{giannozzi2009quantum,
  title={QUANTUM ESPRESSO: a modular and open-source software project for quantum simulations of materials},
  author={Giannozzi, Paolo and Baroni, Stefano and Bonini, Nicola and Calandra, Matteo and Car, Roberto and Cavazzoni, Carlo and Ceresoli, Davide and Chiarotti, Guido L and Cococcioni, Matteo and Dabo, Ismaila and others},
  journal={Journal of physics: Condensed Matter},
  volume={21},
  number={39},
  pages={395502},
  year={2009},
  publisher={IOP Publishing},
  url={https://doi.org/10.1088/0953-8984/21/39/395502}
}

@article{giannozzi2017advanced,
  title={Advanced capabilities for materials modelling with Quantum ESPRESSO},
  author={Giannozzi, Paolo and Andreussi, Oliviero and Brumme, Thomas and Bunau, Oana and Nardelli, M Buongiorno and Calandra, Matteo and Car, Roberto and Cavazzoni, Carlo and Ceresoli, Davide and Cococcioni, Matteo and others},
  journal={Journal of Physics: Condensed Matter},
  volume={29},
  number={46},
  pages={465901},
  year={2017},
  publisher={IOP Publishing},
  url={https://iopscience.iop.org/article/10.1088/1361-648X/aa8f79}
  }

@article{qe_exa_2020,
    author = {Giannozzi, Paolo and Baseggio, Oscar and Bonfà, Pietro and Brunato, Davide and Car, Roberto and Carnimeo, Ivan and Cavazzoni, Carlo and de Gironcoli, Stefano and Delugas, Pietro and Ferrari Ruffino, Fabrizio and Ferretti, Andrea and Marzari, Nicola and Timrov, Iurii and Urru, Andrea and Baroni, Stefano},
    title = {Quantum ESPRESSO toward the exascale},
    journal = {The Journal of Chemical Physics},
    volume = {152},
    number = {15},
    pages = {154105},
    year = {2020},
    month = {04},
    issn = {0021-9606},
    doi = {10.1063/5.0005082},
    url = {https://doi.org/10.1063/5.0005082},
}

@article{perdew1996generalized,
  title={Generalized gradient approximation made simple},
  author={Perdew, John P and Burke, Kieron and Ernzerhof, Matthias},
  journal={Physical Review Letters},
  volume={77},
  number={18},
  pages={3865},
  year={1996},
  publisher={APS},
  url={https://doi.org/10.1103/PhysRevLett.77.3865}
}

@article{baroni1987green,
  title={Green’s-function approach to linear response in solids},
  author={Baroni, Stefano and Giannozzi, Paolo and Testa, Andrea},
  journal={Physical Review Letters},
  volume={58},
  number={18},
  pages={1861},
  year={1987},
  publisher={APS},
  url={https://doi.org/10.1103/PhysRevLett.58.1861}
}

@article{gonze1997first,
  title={First-principles responses of solids to atomic displacements and homogeneous electric fields: Implementation of a conjugate-gradient algorithm},
  author={Gonze, Xavier},
  journal={Physical Review B},
  volume={55},
  number={16},
  pages={10337},
  year={1997},
  publisher={APS},
  url={https://doi.org/10.1103/PhysRevB.55.10337}
  }

@article{lejaeghere2016reproducibility,
  title={Reproducibility in density functional theory calculations of solids},
  author={Lejaeghere, Kurt and Bihlmayer, Gustav and Bj{\"o}rkman, Torbj{\"o}rn and Blaha, Peter and Bl{\"u}gel, Stefan and Blum, Volker and Caliste, Damien and Castelli, Ivano E and Clark, Stewart J and Dal Corso, Andrea and others},
  journal={Science},
  volume={351},
  number={6280},
  pages={aad3000},
  year={2016},
  publisher={American association for the advancement of science},
  url={https://doi.org/10.1126/science.aad3000}
}

@article{prandini2018precision,
  title={Precision and efficiency in solid-state pseudopotential calculations},
  author={Prandini, Gianluca and Marrazzo, Antimo and Castelli, Ivano E and Mounet, Nicolas and Marzari, Nicola},
  journal={npj Computational Materials},
  volume={4},
  number={1},
  pages={72},
  year={2018},
  publisher={Nature Publishing Group UK London},
  url={https://doi.org/10.1038/s41524-018-0127-2}
}

@article{bastonero2025first,
  title={First-principles Hubbard parameters with automated and reproducible workflows},
  author={Bastonero, Lorenzo and Malica, Cristiano and Macke, Eric and Bercx, Marnik and Huber, Sebastiaan and Timrov, Iurii and Marzari, Nicola},
  journal={npj Computational Materials},
  volume={11},
  number={1},
  pages={183},
  year={2025},
  publisher={Nature Publishing Group UK London},
  url={https://doi.org/10.1038/s41524-025-01685-4}
}

@article{timrov2018hubbard,
  title = {Hubbard parameters from density-functional perturbation theory},
  author = {Timrov, Iurii and Marzari, Nicola and Cococcioni, Matteo},
  journal = {Phys. Rev. B},
  volume = {98},
  issue = {8},
  pages = {085127},
  numpages = {15},
  year = {2018},
  month = {Aug},
  publisher = {American Physical Society},
  doi = {10.1103/PhysRevB.98.085127},
  url = {https://link.aps.org/doi/10.1103/PhysRevB.98.085127}
}

@article{timrov2022hp,
  title={HP--A code for the calculation of Hubbard parameters using density-functional perturbation theory},
  author={Timrov, Iurii and Marzari, Nicola and Cococcioni, Matteo},
  journal={Computer Physics Communications},
  doi = {https://doi.org/10.1016/j.cpc.2022.108455},
  url = {https://www.sciencedirect.com/science/article/pii/S0010465522001746},
  volume={279},
  pages={108455},
  year={2022},
  publisher={Elsevier}
}

@Article{10.21468/SciPostPhysCodeb.30,
	title={{A tool to check whether a symmetry-compensated collinear magnetic material is antiferro- or altermagnetic}},
	author={Andriy Smolyanyuk and Libor \v{S}mejkal and Igor I. Mazin},
	journal={SciPost Physics Codebases},
	pages={30},
	year={2024},
	publisher={SciPost},
	doi={10.21468/SciPostPhysCodeb.30},
	url={https://scipost.org/10.21468/SciPostPhysCodeb.30},
}

@Article{10.21468/SciPostPhysCodeb.30-r1.0,
	title={{Codebase release r1.0 for amcheck}},
	author={Andriy Smolyanyuk and Libor \v{S}mejkal and Igor I. Mazin},
	journal={SciPost Physics Codebases},
	pages={30-r1.0},
	year={2024},
	publisher={SciPost},
	doi={10.21468/SciPostPhysCodeb.30-r1.0},
	url={https://scipost.org/10.21468/SciPostPhysCodeb.30-r1.0},
}

@article{yu2014communication,
  title={Communication: Comparing ab initio methods of obtaining effective U parameters for closed-shell materials},
  author={Yu, Kuang and Carter, Emily A},
  journal={The Journal of Chemical Physics},
  volume={140},
  number={12},
  year={2014},
  publisher={AIP Publishing},
  url={https://doi.org/10.1063/1.4869718}
}

@article{de2026accurate,
  title={Accurate and efficient protocols for high-throughput first-principles materials simulations},
  author={de Miranda Nascimento, Gabriel and dos Santos, Flaviano Jos{\'e} and Bercx, Marnik and Grassano, Davide and Pizzi, Giovanni and Marzari, Nicola},
  journal={npj Computational Materials},
  year={2026},
  volume={},
  number={},
  publisher={Nature Publishing Group UK London},
  url={https://doi.org/10.1038/s41524-026-02097-8}
}

@article{liechtenstein1995density,
  title={Density-functional theory and strong interactions: Orbital ordering in {Mott-Hubbard} insulators},
  author={Liechtenstein, AI and Anisimov, Vladimir I and Zaanen, Jan},
  journal={Physical Review B},
  volume={52},
  number={8},
  pages={R5467},
  year={1995},
  publisher={APS},
  url={https://doi.org/10.1103/PhysRevB.52.R5467}
}

@article{cococcioni2005linear,
  title={Linear response approach to the calculation of the effective interaction parameters in the {LDA+U} method},
  author={Cococcioni, Matteo and De Gironcoli, Stefano},
  journal={Physical Review B},
  volume={71},
  number={3},
  pages={035105},
  year={2005},
  publisher={APS},
  url={ https://doi.org/10.1103/PhysRevB.71.035105}
}

@article{spaldin2026there,
  title={Why are there so few non-altermagnetic antiferromagnets?},
  author={Spaldin, Nicola A and Cheong, Sang-Wook and Griffin, Sinead},
  journal={arXiv preprint arXiv:2602.17181},
  url={https://doi.org/10.48550/arXiv.2602.17181},
  year={2026}
}

@article{yu2026identifying,
  title={Identifying Oriented Spin Space Groups and Related Physical Properties Using an Online Platform FINDSPINGROUP},
  author={Yu, Yutong and Chen, Xiaobing and Zhu, Yanzhou and Li, Yuhui and Xiong, Renzheng and Li, Jiayu and Liu, Yuntian and Liu, Qihang},
  journal={arXiv preprint arXiv:2604.21397},
  url={https://doi.org/10.48550/arXiv.2604.21397},
  year={2026}
}

@article{huber2020aiida,
  title={AiiDA 1.0, a scalable computational infrastructure for automated reproducible workflows and data provenance},
  author={Huber, Sebastiaan P and Zoupanos, Spyros and Uhrin, Martin and Talirz, Leopold and Kahle, Leonid and H{\"a}uselmann, Rico and Gresch, Dominik and M{\"u}ller, Tiziano and Yakutovich, Aliaksandr V and Andersen, Casper W and others},
  journal={Scientific Data},
  volume={7},
  number={1},
  pages={300},
  year={2020},
  publisher={Nature Publishing Group UK London},
  url={https://doi.org/10.1038/s41597-020-00638-4}
}

@article{uhrin2021workflows,
  title={Workflows in AiiDA: Engineering a high-throughput, event-based engine for robust and modular computational workflows},
  author={Uhrin, Martin and Huber, Sebastiaan P and Yu, Jusong and Marzari, Nicola and Pizzi, Giovanni},
  journal={Computational Materials Science},
  volume={187},
  pages={110086},
  year={2021},
  publisher={Elsevier},
  doi = {https://doi.org/10.1016/j.commatsci.2020.110086},
url = {https://www.sciencedirect.com/science/article/pii/S0927025620305772},
}

@article{methfessel1989high,
  title={High-precision sampling for Brillouin-zone integration in metals},
  author={Methfessel, MPAT and Paxton, Anthony T},
  journal={Physical Review B},
  volume={40},
  number={6},
  pages={3616},
  year={1989},
  publisher={APS},
  url={https://doi.org/10.1103/PhysRevB.40.3616}
}

@article{marzari1999thermal,
  title={Thermal contraction and disordering of the Al (110) surface},
  author={Marzari, Nicola and Vanderbilt, David and De Vita, Alessandro and Payne, Mike C},
  journal={Physical Review Letters},
  volume={82},
  number={16},
  pages={3296},
  year={1999},
  publisher={APS},
  url={https://doi.org/10.1103/PhysRevLett.82.3296}
}

@article{vanderbilt1990soft,
  title={Soft self-consistent pseudopotentials in a generalized eigenvalue formalism},
  author={Vanderbilt, David},
  journal={Physical Review B},
  volume={41},
  number={11},
  pages={7892},
  year={1990},
  publisher={APS},
  url={http://journals.aps.org/prb/abstract/10.1103/PhysRevB.41.7892}
}

@article{Blochl1994projector,
  title = {Projector augmented-wave method},
  author = {Bl\"ochl, P. E.},
  journal = {Physical Review B},
  volume = {50},
  issue = {24},
  pages = {17953--17979},
  numpages = {0},
  year = {1994},
  month = {Dec},
  publisher = {American Physical Society},
  doi = {10.1103/PhysRevB.50.17953},
  url = {https://link.aps.org/doi/10.1103/PhysRevB.50.17953}
}

@article{Hamann2013optimized,
  title = {Optimized norm-conserving Vanderbilt pseudopotentials},
  author = {Hamann, D. R.},
  journal = {Physical Review B},
  volume = {88},
  issue = {8},
  pages = {085117},
  numpages = {10},
  year = {2013},
  month = {Aug},
  publisher = {American Physical Society},
  doi = {10.1103/PhysRevB.88.085117},
  url = {https://link.aps.org/doi/10.1103/PhysRevB.88.085117}
}

@article{Goedecker1996separable,
  title = {Separable dual-space Gaussian pseudopotentials},
  author = {Goedecker, S. and Teter, M. and Hutter, J.},
  journal = {Physical Review B},
  volume = {54},
  issue = {3},
  pages = {1703--1710},
  numpages = {0},
  year = {1996},
  month = {Jul},
  publisher = {American Physical Society},
  doi = {10.1103/PhysRevB.54.1703},
  url = {https://link.aps.org/doi/10.1103/PhysRevB.54.1703}
}

@article{willand2013norm,
  title={Norm-conserving pseudopotentials with chemical accuracy compared to all-electron calculations},
  author={Willand, Alex and Kvashnin, Yaroslav O and Genovese, Luigi and V{\'a}zquez-Mayagoitia, {\'A}lvaro and Deb, Arpan Krishna and Sadeghi, Ali and Deutsch, Thierry and Goedecker, Stefan},
  journal={The Journal of Chemical Physics},
  volume={138},
  number={10},
  year={2013},
  publisher={AIP Publishing},
  url={http://dx.doi.org/10.1063/1.4793260}
}

@article{kucukbenli2014projector,
  title={Projector augmented-wave and all-electron calculations across the periodic table: a comparison of structural and energetic properties},
  author={Kucukbenli, E and Monni, M and Adetunji, BI and Ge, X and Adebayo, GA and Marzari, N and De Gironcoli, S and Corso, A Dal},
  journal={arXiv preprint arXiv:1404.3015},
  year=2014,
  url={https://arxiv.org/abs/1404.3015}
}

@article{dal2014pseudopotentials,
  title={Pseudopotentials periodic table: From H to Pu},
  author={Dal Corso, Andrea},
  journal={Computational Materials Science},
  volume={95},
  pages={337--350},
  year={2014},
  publisher={Elsevier},
  url={https://doi.org/10.1016/j.commatsci.2014.07.043}
}

@article{garrity2014pseudopotentials,
  title={Pseudopotentials for high-throughput DFT calculations},
  author={Garrity, Kevin F and Bennett, Joseph W and Rabe, Karin M and Vanderbilt, David},
  journal={Computational Materials Science},
  volume={81},
  pages={446--452},
  year={2014},
  publisher={Elsevier},
  url={http://dx.doi.org/10.1016/j.commatsci.2013.08.053}
}

@article{schlipf2015optimization,
  title={Optimization algorithm for the generation of ONCV pseudopotentials},
  author={Schlipf, Martin and Gygi, Fran{\c{c}}ois},
  journal={Computer Physics Communications},
  volume={196},
  pages={36--44},
  year={2015},
  publisher={Elsevier},
  url={http://dx.doi.org/10.1016/j.cpc.2015.05.011}
}

@article{he2013phase,
  title={Phase diagram and electronic indication of high-temperature superconductivity at 65 K in single-layer {FeSe} films},
  author={He, Shaolong and He, Junfeng and Zhang, Wenhao and Zhao, Lin and Liu, Defa and Liu, Xu and Mou, Daixiang and Ou, Yun-Bo and Wang, Qing-Yan and Li, Zhi and others},
  journal={Nature Materials},
  volume={12},
  number={7},
  pages={605--610},
  year={2013},
  publisher={Nature Publishing Group UK London},
  url={https://doi.org/10.1038/nmat3648}
}

@article{lee2014interfacial,
  title={Interfacial mode coupling as the origin of the enhancement of Tc in {FeSe} films on {SrTiO$_3$}},
  author={Lee, JJ and Schmitt, FT and Moore, RG and Johnston, S and Cui, Y-T and Li, W and Yi, M and Liu, ZK and Hashimoto, M and Zhang, Ya and others},
  journal={Nature},
  volume={515},
  number={7526},
  pages={245--248},
  year={2014},
  publisher={Nature Publishing Group UK London},
  url={https://doi.org/10.1038/nature13894}
}

@article{torelli2020high,
  title={High-throughput computational screening for two-dimensional magnetic materials based on experimental databases of three-dimensional compounds},
  author={Torelli, Daniele and Moustafa, Hadeel and Jacobsen, Karsten W and Olsen, Thomas},
  journal={npj Computational Materials},
  volume={6},
  number={1},
  pages={158},
  year={2020},
  publisher={Nature Publishing Group UK London},
  url={https://doi.org/10.1038/s41524-020-00428-x}
}

@article{zeng2026classification,
title={Classification and design of two-dimensional altermagnets},
author={Zeng, Sike and Liu, Dong and Peng, Hongjie and He, Chang-Chun and Yang, Xiao-Bao and Zhao, Yu-Jun},
journal = {Frontiers of Physics},
volume = {21},
pages = {095301-},
year = {2026},
issn = {2095-0462},
doi = {https://doi.org/10.15302/frontphys.2026.095301},
url = {https://journal.hep.com.cn/fop/EN/10.15302/frontphys.2026.095301}
}

@article{xu2026chemical,
  title={Chemical design of monolayer altermagnets},
  author={Xu, Runzhang and Gao, Yifan and Liu, Junwei},
  journal={National Science Review},
  volume={13},
  number={2},
  pages={nwaf528},
  year={2026},
  url={https://doi.org/10.1093/nsr/nwaf528},
  publisher={Oxford University Press}
}

@article{peng2025multicomponent,
  title = {Multicomponent altermagnet: A general approach to generating multicomponent structures with two-dimensional altermagnetism},
  author = {Peng, Hongjie and Zeng, Sike and Liao, Ji-Hai and He, Chang-Chun and Yang, Xiao-Bao and Zhao, Yu-Jun},
  journal = {Physical Review B},
  volume = {111},
  issue = {19},
  pages = {195123},
  numpages = {12},
  year = {2025},
  month = {May},
  publisher = {American Physical Society},
  doi = {10.1103/PhysRevB.111.195123},
  url = {https://link.aps.org/doi/10.1103/PhysRevB.111.195123}
}

@article{mak2019probing,
  title={Probing and controlling magnetic states in 2D layered magnetic materials},
  author={Mak, Kin Fai and Shan, Jie and Ralph, Daniel C},
  journal={Nature Reviews Physics},
  volume={1},
  number={11},
  pages={646--661},
  year={2019},
  publisher={Nature Publishing Group UK London},
  url={https://doi.org/10.1038/s42254-019-0110-y}
}

@article{ma2021multifunctional,
  title={Multifunctional antiferromagnetic materials with giant piezomagnetism and noncollinear spin current},
  author={Ma, Hai-Yang and Hu, Mengli and Li, Nana and Liu, Jianpeng and Yao, Wang and Jia, Jin-Feng and Liu, Junwei},
  journal={Nature communications},
  volume={12},
  number={1},
  pages={2846},
  year={2021},
  publisher={Nature Publishing Group UK London},
  url={https://doi.org/10.1038/s41467-021-23127-7}
}

@article{jiang2025metallic,
  title={A metallic room-temperature d-wave altermagnet},
  author={Jiang, Bei and Hu, Mingzhe and Bai, Jianli and Song, Ziyin and Mu, Chao and Qu, Gexing and Li, Wan and Zhu, Wenliang and Pi, Hanqi and Wei, Zhongxu and others},
  journal={Nature Physics},
  volume={21},
  number={5},
  pages={754--759},
  year={2025},
  publisher={Nature Publishing Group UK London},
  url={https://doi.org/10.1038/s41567-025-02822-y}
}

@article{zhang2025crystal,
  title={Crystal-symmetry-paired spin--valley locking in a layered room-temperature metallic altermagnet candidate},
  author={Zhang, Fayuan and Cheng, Xingkai and Yin, Zhouyi and Liu, Changchao and Deng, Liwei and Qiao, Yuxi and Shi, Zheng and Zhang, Shuxuan and Lin, Junhao and Liu, Zhengtai and others},
  journal={Nature Physics},
  volume={21},
  number={5},
  pages={760--767},
  year={2025},
  publisher={Nature Publishing Group UK London},
  url={https://doi.org/10.1038/s41567-025-02864-2}
}

@article{spglib,
  author = {Atsushi Togo, Kohei Shinohara and Isao Tanaka},
  title = {Spglib: a software library for crystal symmetry search},
  journal = {Science and Technology of Advanced Materials: Methods},
  volume = {4},
  number = {1},
  pages = {2384822--2384836},
  year = {2024},
  doi = {10.1080/27660400.2024.2384822},
  url = {https://doi.org/10.1080/27660400.2024.2384822},
}

@article{spglibv2,
  author = {Shinohara, Kohei and Togo, Atsushi and Tanaka, Isao},
  title = {Algorithms for magnetic symmetry operation search and identification of magnetic space group from magnetic crystal structure},
  journal = {Acta Crystallographica Section A},
  year = {2023},
  volume = {79},
  number = {5},
  pages = {390--398},
  month = {Sep},
  doi = {10.1107/S2053273323005016},
  url = {https://doi.org/10.1107/S2053273323005016},
}





\end{document}